%% file: main.tex
\documentclass[conference,compsoc]{IEEEtran}
\ifCLASSOPTIONcompsoc
  \usepackage[nocompress]{cite}
\else
  \usepackage{cite}
\fi
\ifCLASSINFOpdf
  \usepackage[pdftex]{graphicx}
  \graphicspath{{./images/}{./figures/}}
  \DeclareGraphicsExtensions{.pdf,.jpeg,.jpg,.png}
\else
  \usepackage[dvips]{graphicx}
  \graphicspath{{./images/}{./figures/}}
  \DeclareGraphicsExtensions{.eps}
\fi
\usepackage{mdframed}

\usepackage[hyphens]{url}

\usepackage{balance}

\usepackage{booktabs}
\newcommand{\ra}[1]{\renewcommand{\arraystretch}{#1}}

\usepackage{multirow}

\input{setup-comments}

\input{setup-massivizing}

\begin{document}
%
\title{Massivizing Computer Systems: a Vision to Understand, Design, and Engineer Computer Ecosystems through and beyond Modern Distributed Systems}

\author{\IEEEauthorblockN{Alexandru Iosup}
\IEEEauthorblockA{Department of Computer Science,\\
Faculty of Sciences, VU Amsterdam,\\
The Netherlands\\
A.Iosup@vu.nl}
\and
\IEEEauthorblockN{Alexandru Uta}
\IEEEauthorblockA{Department of Computer Science,\\
Faculty of Sciences, VU Amsterdam,\\
The Netherlands\\
A.Uta@vu.nl}
\and
\IEEEauthorblockN{The AtLarge Team\IEEEauthorrefmark{1}}
\IEEEauthorblockA{VU Amsterdam and\\
Delft University of Technology,\\
The Netherlands,\\
\url{https://atlarge-research.com}}%
\IEEEcompsocitemizethanks{
\IEEEcompsocthanksitem The AtLarge team members co-authoring this article are: Georgios Andreadis, Vincent van Beek, Erwin van Eyk, Tim Hegeman,  Sacheendra Talluri, Lucian Toader, and Laurens Versluis.}%
}


%


\maketitle

\input{content-abstract}

\input{content-intro}

\input{content-main-ecosystems}

\input{content-main-core}

\input{content-main-challenges}

\input{content-main-usecases}

\input{content-main-roadmap}
\input{content-related}

\input{content-conclusion}

{\balance
\section*{Acknowledgments}
{\small
This work is supported 
by the Dutch projects Vidi MagnaData and KIEM KIESA, 
by the Dutch Commit and the Commit project Commissioner, and 
by generous donations from Oracle Labs, USA.
We thank our first (internal) reviewers at the Delft University of Technology and Vrije Universiteit Amsterdam, the Netherlands, including Jan Rellermeyer and Cristiano Giuffrida, respectively.
We are indebted to Dick H. J. Epema, whose ideas have helped shape the career of the first author and thus this vision of the AtLarge research group.
We thank for discussing ideas related to this work to former and current members of the AtLarge team\footnote{\url{https://atlarge-research.com/people.html}}, to our collaborators from the SPEC RG Cloud group, and to our collaborators from the LDBC consortium, including: 
Cristina L. Abad (Escuela Superior Polit\'{e}cnica del Litoral, Ecuador);
Tarek Abdelzaher (University of Illinois at Urbana Champaign, IL, USA);
Andre Bauer, Simon Eismann, Johannes Grohmann, Nikolas Herbst, Samuel Kounev, and Simon Spinner (U. Wuertzburg, Germany); 
Sara Bouchenak (INSA Lyon, France);
Tim Brecht (University of Waterloo, Canada);
Mihai Capot\~{a} (Intel Labs, Portland, OR, USA);
Hassan Chafi, Sungpack Hong, Thomas Manhardt, and Petr Koupy (Oracle Labs, SFBA, USA);
Ahmed Ali El-Din (UMASS, USA); 
Athanasia Evangelinou and George Kousiouris (Nat'l. Tech. U. of Athens, Greece); 
Bogdan Ghit and Wing Lung Ngai (Databricks Amsterdam, the Netherlands);
Stijn Heldens (U. Twente, the Netherlands);
Eva Kalyvianaki (Imperial College of London, UK);
Rouven Krebs and Simon Seif (SAP SE, Germany);
Martina Maggio (Lund University, Sweden);
Ole Mengshoel (CMU Silicon Valley at the NASA Ames Research Center, PA, USA);
Arif Merchant (Google, Inc., CA, USA);
Alessandro V. Papadopoulos (Maelardalen University, Sweden);
Arnau Prat P\'{e}rez and Josep Larriba Pey (UPC Barcelona);
Andy Tanenbaum and Peter Boncz (Vrije Universiteit Amsterdam, the Netherlands);
Ilie Gabriel T\u{a}nase (IBM TJ Watson Research Center, NY, USA);
Markus Th\"{o}mmes (IBM, Germany);
Yinglong Xia (Huawei Munich, Germany); and
Xiaoyun Zhu (Futurewei Technologies, CA, USA).
}
\bibliographystyle{IEEEtran}
\bibliography{IEEEabrv,main_beautified}
}

\end{document}

%% file: setup-comments.tex
\usepackage{xcolor}  
\definecolor{OwnAzure}{HTML}{336699}
\definecolor{OwnCerulean}{HTML}{CAE2FE}
\definecolor{OwnOliveGreen}{HTML}{556B2F}
\usepackage[colorinlistoftodos,prependcaption,textsize=tiny]{todonotes}
\usepackage{xargs}                      
\newcommandx{\todolarge}[2][1=]{\todo[inline,size=\large,linecolor=OwnAzure,backgroundcolor=OwnCerulean,bordercolor=OwnAzure,#1]{#2}}
\newcommandx{\todoai}[2][1=]{\todo[inline,linecolor=OwnAzure,backgroundcolor=OwnCerulean,bordercolor=OwnAzure,#1]{#2}}

\newcommandx{\addref}[2][1=]
{\todo[inline,linecolor=blue,backgroundcolor=blue!50,bordercolor=blue,#1]{Add reference. #2}}
\newcommandx{\unsure}[2][1=]{\todo[inline, linecolor=red,backgroundcolor=red!25,bordercolor=red,#1]{#2}}
\newcommandx{\change}[2][1=]{\todo[inline, linecolor=blue,backgroundcolor=blue!25,bordercolor=blue,#1]{#2}}
\newcommandx{\info}[2][1=]{\todo[linecolor=OwnOliveGreen,backgroundcolor=OwnOliveGreen!25,bordercolor=OwnOliveGreen,#1]{#2}}
\newcommandx{\improvement}[2][1=]{\todo[linecolor=Plum,backgroundcolor=Plum!25,bordercolor=Plum,#1]{#2}}
\newcommandx{\thiswillnotshow}[2][1=]{\todo[disable,#1]{#2}}

\definecolor{darkred}{rgb}{0.5,0,0}
\definecolor{darkgreen}{rgb}{0,0.5,0}
\definecolor{darkblue}{rgb}{0,0,0.5}



%% file: setup-massivizing.tex
\newcommand\ourdomain{Massivizing Computer Systems}
\newcommand\ourdomainshort{{\sc MCS}}

\newcommand\DistribSys{Distributed Systems}

\newcommand\CompSys{Computer Systems}

\newcommand\SwEng{Software Engineering}

\newcommand\PerfEng{Performance Engineering}

\newcommand\BioSys{Systems Biology}

\usepackage{tcolorbox}
\tcbuselibrary{skins}

\definecolor{BgBlue}{HTML}{336699}
\definecolor{BgGreen}{HTML}{339966}
\definecolor{BgYellow}{HTML}{664422}
\definecolor{BgOrange}{HTML}{994422}

\newcommand\vision[1]{
\begin{tcolorbox}[colback=BgYellow!15, colframe=BgYellow!90!black,enhanced,drop fuzzy shadow,sharp corners]
\textbf{Vision}: #1
\end{tcolorbox}
}

\newcommand\definition[1]{
\begin{tcolorbox}[colback=orange!15, colframe=orange!90!black,enhanced,drop fuzzy shadow,sharp corners]
\textbf{Definition}: #1
\end{tcolorbox}
}

\newcommand\namedbox[2]{
\begin{tcolorbox}[colback=orange!15, colframe=orange!90!black,enhanced,drop fuzzy shadow,sharp corners]
\textbf{#1}: #2
\end{tcolorbox}
}

\newcounter{principleid}
\newcommand{\rprinciple}[1]{\refstepcounter{principleid}\label{#1}}

\newcommand\principle[2]{
\rprinciple{#1}
\begin{tcolorbox}[colback=BgBlue!15, colframe=BgBlue!90!black,enhanced,drop fuzzy shadow,sharp corners]
\textbf{P\arabic{principleid}}: #2
\end{tcolorbox}
}

\newcommand\refpr[2]{\textbf{P\ref{#1}}}

\newcounter{challengeid}
\newcommand{\rchallenge}[1]{\refstepcounter{challengeid}\label{#1}}

\newcommand\challenge[3]{ 
\rchallenge{#1}
\begin{tcolorbox}[colback=BgGreen!15, colframe=BgGreen!90!black,enhanced,drop fuzzy shadow,sharp corners]
\textbf{C\arabic{challengeid}}: #3 (From #2.)
\end{tcolorbox}
}

\usepackage{mathabx}

%% file: content-abstract.tex
\begin{abstract}

Our society is digital: industry, science, governance, and individuals depend, often transparently, on the inter-operation of large numbers of distributed computer systems. 

Although the society takes them almost for granted, these computer ecosystems are not available for all, may not be affordable for long, and raise numerous other research challenges. 

Inspired by these challenges and by our experience with distributed computer systems, 
we envision Massivizing Computer Systems, a domain of computer science
focusing on understanding, controlling, and evolving successfully such ecosystems. 
Beyond establishing and growing a body of knowledge about computer ecosystems and their constituent systems, the community in this domain should also aim to educate many about design and engineering for this domain, and all people about its principles. 
This is a call to the entire community: there is
much to discover and achieve.

\end{abstract}

%% file: content-intro.tex
\section{Introduction}
\label{sec:introduction}\label{sec:intro}

The modern lifestyle depends on computer\footnote{The analysis by E.W. Dijkstra~\cite{DBLP:conf/sac/Dijkstra99} explains the main differences between ``computer'' and ``computing'' science: origins in the US vs. in the (current) EU, respectively, with American CS seen in the past as ``more machine-oriented, less mathematical, more closely linked to application areas, more quantitative, and more willing to absorb industrial products in its curriculum''. The differences have now softened, and participants beyond US and EU have since joined our community.} {\it ecosystems}. 
We engage increasingly with each other, with governance, and with the Digital Economy~\cite{tapscott1996digital} through diverse computer ecosystems comprised of globally distributed systems, 
developed and operated by diverse organizations,
interoperated across diverse legal and administrative boundaries. 
These computer ecosystems create economies of scale, and underpin participation and innovation in the knowledge-based society: 
for example, in the European Union, 
information and communication technology~(ICT)\footnote{ICT loosely encompasses all technology and processes used to process information (the ``I'') and for communications (the ``C''). Historically, the distinction between the ``I'' and the ``C'' in ICT can be traced to the early days of computing, where information was stored and processed as {\it digital} data, and most communication was based on {\it analog} devices. This distinction has lost importance starting with the advent of all-digital networks, completed in the 1990s Internet.}, for which all services are migrating to computer
 ecosystems\footnote{Computer ecosystems build a world of 
cloud computing~\cite{mell2011nist}, artificial intelligence~\cite{DBLP:journals/nature/SilverHMGSDSAPL16}, and big data~\cite{john2014big}, underpinned by 
diverse software systems and networks interconnecting datacenters~\cite{DBLP:conf/sigcomm/RoyZBPS15}, and edge~\cite{DBLP:journals/ccr/LopezMEDHIBFR15}/smart devices.}, 
accounts for nearly 5\% of the economy and accounts for nearly 50\% of productivity growth\footnote{Correspondingly, ICT receives about 25\% of all business R\&D funding and is at the core of EU's H2020 programme, see \url{https://ec.europa.eu/programmes/horizon2020/en/area/ict-research-innovation}.}. 
However positive, computer ecosystems are not merely larger, deeper structures (e.g., hierarchies) of distributed computer systems. Although we have conquered many of the scientific and engineering challenges of distributed computer systems\footnote{M. van Steen and A. Tanenbaum provide an introduction~\cite{DBLP:journals/computing/SteenT16}.},
computer ecosystems add numerous challenges stemming from the complexity of structure, organization, and evolving and emerging use.
We envision in this work how computer systems can further develop as a positive technology for our society. 

\vision{
We envision a world where individuals and human-centered organizations are augmented by an automated, sustainable layer of technology. At the core of this technology is ICT, and at the core of ICT are computer ecosystems, interoperating and performing as utilities and services, under human guidance and control. 
In our vision, ICT is a fundamental human right, including the right to learn how to use this technology. 
}

We see a fundamental crisis, the {\it ecosystems crisis}, already at work and hampering our vision. 
The natural evolution from early \CompSys{} to modern \DistribSys{} has been until now halted by relatively few crises, among which standing out is the software crisis of the 1960s, due to unbounded increase in complexity~\cite{DBLP:journals/cacm/Dijkstra72,DBLP:journals/annals/Wirth08}. 
We see the ongoing ecosystems crisis as due to similar reasons, and leading the \DistribSys{} field to a fundamental deficit of knowledge and of technology\footnote{Like Arthur~\cite[Ch.2]{design:book/Arthur09}, we refute the dictionary definition of technology, which superficially places technology in a role secondary to (applied) science. Instead, we use the first-principle definition provided by Arthur: technology is (i) use-driven, (ii) a group of practices and components, typically becoming useful through the execution of a sequence of operations, (iii) the set of groups from (iv) across all engineering available to a human culture, forming thus Kevin Kelly's ``technium''.}, with abundant forewarnings. 
In Section~\ref{sec:bg:ecosystems}, we define and give practical examples of five fundamental problems of computer ecosystems that we believe apply even to the most successful of the tech companies, such as Amazon, Alibaba, Google, Facebook, etc.
but even more so to the small and medium enterprises that should develop the next generation of technology: 
(i) lacking the core laws and theories of computer ecosystems;
(ii) lacking the technology to maintain today's computer ecosystems;
(iii) lacking the instruments to design, tune, and operate computer ecosystems against foreseeable needs;
(iv) lacking peopleware\footnote{``If the organization is a development shop, it will optimize for the short term, exploit people, cheat on the workplace, and do nothing to conserve its very lifeblood, the peopleware that is its only real asset. If we ran our agricultural economy on the same basis, we'd eat our seed corn immediately and starve next year.''~\cite[Kindle Loc. 1482-1484]{concepts:book/DeMarcoL12}.} knowledge and processes; and
(v) going beyond mere technology.

Our vision, of {\it Massivizing Computer Systems}, 
focuses on rethinking the body of knowledge, and the peopleware and methodological processes, associated with computer ecosystems.
We aim to reuse what is valuable and available in \DistribSys{}, and in the complementary fields of \SwEng{} and \PerfEng{}, and to further develop only what is needed. 
Grid computing and cloud computing, which both leverage the advent of the Networked World\footnote{Of which the Internet is a prominent example, but which further includes networking in supercomputing, telco, and IoT-focused industries.}, of modern processes for the design and development of software systems, and of modern techniques for performance engineering, are sources of technology for utility computing\footnote{We trace the use of ``utility computing'' in scientific publications to Andrzejak, Arlitt, and Rolia~\cite{misc:AndrzejakAR02}, and to Buyya~\cite{DBLP:conf/cluster/Buyya03}.}. 
However, grid computing has succumbed to the enormous complexity of the ecosystems crisis, for example, it did not reach needed automation for heterogeneous resources and non-functional requirements such as elasticity, and did not develop appropriate cost models.

Armed with knowledge and practical tools similar to grid computing, the pragmatic and economically viable \DistribSys{} domain of cloud computing started with the limited goal of building a digital ecosystem where the core is largely homogeneous, and is still primarily operated from single-organization datacenters. Attempts to expand to more diverse ecosystems have led to problems, some of which we have already covered.
Edge-centric computing~\cite{DBLP:journals/ccr/LopezMEDHIBFR15} borrows from peer-to-peer computing and proposes to shift control to nodes at the edge, closer to the user and thus human-centric in its security and trust models, but still relies on current cloud technology instead of explicitly managing the full-stack complexity of ecosystems.

We propose to complement and extend the existing body of knowledge  with a focus on \ourdomain{}, with the goal of defining and supporting the core body of knowledge and the skills relevant to this vision. 
(This path is successfully followed by other sciences with significant impact in the modern society, such as physics and its impact on high-precision industry, biology and its impact on healthcare, ecology and its impact on wellbeing, etc.)
Toward this goal, we make a five-fold contribution:

\begin{enumerate}
	\item We propose the premises of a new field\footnote{As conjectured by Denning~\cite{DBLP:journals/cacm/Denning13a}, there is a high threshold for becoming a field of science, paraphrasing: focus on the natural and artificial processes of a pervasive phenomenon, a body of knowledge and skills that can be codified and taught, experimental methods of discovery and validation, reproducibility of results and falsifiability of theoretical constructs, the presence of meaningful discovery itself. Even if \ourdomainshort{} does not pass this threshold, the process of exploring it as a new field can lead to surprising discoveries, as in other sciences (see Section~\ref{sec:related}).} of science, design, and engineering focusing on \ourdomainshort{}~(in Section~\ref{sec:core}). To mark this relationship with the vision, we also call the field \ourdomainshort{}. We define \ourdomainshort{} as a part of the \DistribSys{} domain, but also as synthesizing methods from \SwEng{} and \PerfEng{}.

     \item We propose ten core principles (Section~\ref{sec:core:principles}).
    \ourdomainshort{} has not only a technology focus, but also considers peopleware and co-involvement of other sciences. 
    One of the principles has as corollary the periodic revision of principles, and \ourdomainshort{} will apply it---a community challenge.
    
    \item We express the current systems, peopleware, and methodological challenges raised by the field of \ourdomainshort{}~(in Section~\ref{sec:challenges}). We cover diverse topics of research that evolve naturally from ongoing community research in \DistribSys{}, \SwEng{}, and \PerfEng{}. We also raise challenges in the process of designing ecosystems and their constituent systems. 
    
    \item We predict the benefits \ourdomainshort{} can provide to a  set of pragmatic yet high-reward application domains~(in Section~\ref{sec:use}). 
    Overall, we envision that computer ecosystems built on sound principles will lead to significant benefits, such as economies of scale, better non-functional properties of systems,
    lowering the barrier of expertise needed for use, etc.
We consider as immediate application areas
big and democratized (e-)science, 
the future of online gaming and virtual reality,
the future of banking, 
datacenter-based operations including for hosting business-critical workloads,
and serverless app development and operation.

	\item We compare \ourdomainshort{} with other paradigms (Section~\ref{sec:related}). We explicitly compare \ourdomainshort{} with the paradigms emerging from \DistribSys{}, including grid, cloud, and edge-centric computing. 
    We further compare \ourdomainshort{} with paradigms across other sciences and technical sciences.
\end{enumerate}

%% file: content-main-ecosystems.tex
\section{The Problem of Computer Ecosystems}
\label{sec:bg:ecosystems}

In this section, we introduce systems, ecosystems, and five fundamental problems of computer ecosystems.

\subsection{What Are Systems and Ecosystems?}
\label{sec:core:ecosystems}

We use Meadows' definition of systems~\cite[p.188]{systems:book/Meadows08}:
\definition{A system is 
``a set of elements or parts coherently organized and interconnected in a pattern or structure that produces a characteristic set of behaviors, often classified as its ``function'' or ``purpose.'' 
}

The system elements or parts can be systems themselves, producing more fine-grained functions. We see {\it computer ecosystems} as more than just complex computer systems, in that they interact with people and have {\it structure} that is more advanced, combinatorial and hierarchical as is the general nature of technology~\cite{design:book/Arthur09}, etc.:

\definition{
A computer ecosystem is a heterogeneous group of computer systems and, recursively, of computer ecosystems, collectively {\it constituents}. Constituents are autonomous, even in competition with each other. The ecosystem {\it structure} and {\it organization} ensure its collective responsibility: completing functions with humans in the loop, providing desirable non-functional properties that go beyond traditional performance, subject to agreements with clients. 
Ecosystems experience short- and long-term {\it dynamics}: operating well despite challenging, possibly changing conditions external to the control of the ecosystem. 
}

{\bf Collective Responsibility:}
The ecosystem is designed to respond to functional and non-functional requirements.
The ecosystem constituents must be able to act independently of each other, but when they act collectively they can perform {\it collective functions} that are required and that are not possible for any individual system, and/or they can add useful non-functional characteristics to how they perform functions that could still be possible otherwise.
At least some of the collective functions involve the collaboration of a significant fraction of the ecosystem constituents.

{\bf Beyond Performance:}
When collaborating, the ecosystem constituents optimize or satisfice a decision problem focusing on the trade-off between subsets of both the functional and the non-functional requirements, e.g., correct functional result and high performance vs. cost and availability.
The non-functional requirements are diverse, beyond traditional performance: e.g., high performance, high availability and/or reliability, high scalability and/or elasticity, trustworthy and/or secure operation.

{\bf Autonomy:}
ecosystem constituents can often operate autonomously if allowed, and may be self-aware as defined by Kounev et al.~\cite[Def.1.1]{DBLP:books/sp/17/KounevLBBCDEGGGIKZ17}: they could continuously ``learn models capturing knowledge about themselves and the environment'', ``reason using the models [...] enabling them to act [...] in accordance with higher-level goals, which may also be subject to change.''

{\bf How do ecosystems appear?}
Computer ecosystems appear naturally\footnote{Similar ecosystems appear in many areas of technology~\cite[Ch.2,7--9]{design:book/Arthur09}, and in many other kinds of systems~\cite[Ch.5--8]{book/Simon96}.}, through a process of evolution that involves accumulation of technological artifacts in inter-communicating assemblies and hierarchies, and solving increasingly more sophisticated problems\footnote{The co-evolution of problems and solutions appears in all areas of design~\cite{design:book/Lawson05,design:book/Cross11}~\cite[Ch.I, S2--5]{design:book/Brooks10}.}. 
Real-world ecosystems are distributed or include distributed systems among their constituents~\cite{DBLP:journals/computing/SteenT16}, and are operated by and for multiple (competitive) stakeholders. Components often are heterogeneous, built by multiple developers, not using a verified reference architecture, and having to fit with one another despite not being designed end-to-end.

\begin{figure}[!t]
 \centering
 \vspace*{-1cm}
 \includegraphics[width=1.0\columnwidth]{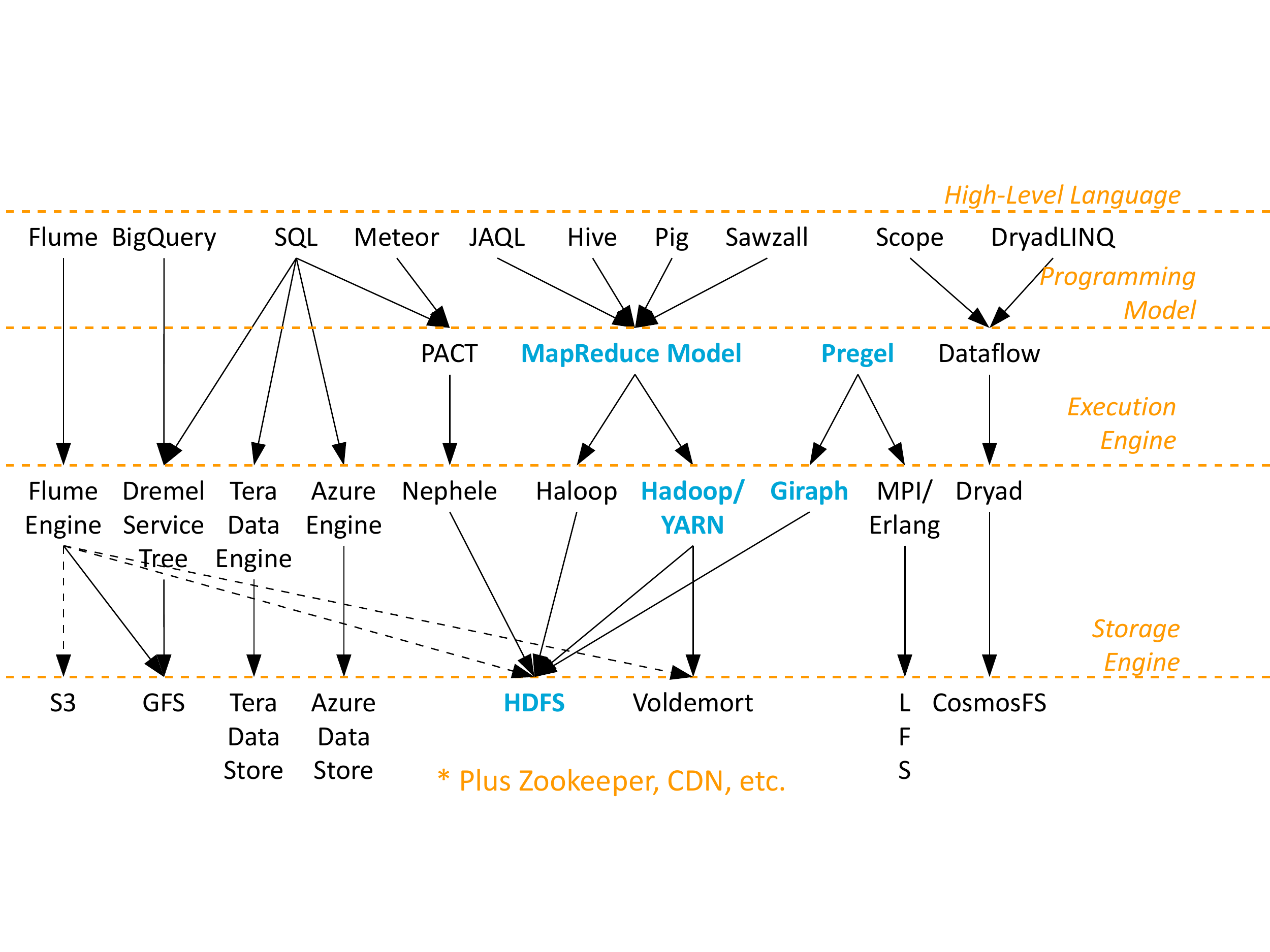} 
 \vspace*{-1.5cm}
 \caption{A view into the ecosystem of Big Data processing. (Reproduced and adapted from our previous work~\cite{DBLP:conf/ccgrid/GhitCHHEI14}.) The four layers, {\it High-Level Language}, {\it Programming Model}, {\it Execution Engine}, and {\it Storage Engine}, are conceptual, but applications that run in this ecosystem typically use components across the full stack of layers (and more, as indicated by the $\star$). The highlighted components cover the minimum set of layers necessary for execution for the MapReduce and Pregel sub-ecosystems.}
 \label{fig:bigdata:ecosystem}
\end{figure}

{\bf A simplified example of ecosystems, sub-ecosystems, and their constituents:}
Developing applications, and tunning, swapping, and adding or removing components requires a deep understanding of the ecosystem.
Figure~\ref{fig:bigdata:ecosystem} depicts the four-layer reference architecture of the big data ecosystem frequently used by the community. 
In this ecosystem, the programming model, e.g., MapReduce, or the execution engine, e.g., Hadoop, typically give name to an entire family of applications of the ecosystem, i.e., ``We run Hadoop applications.'' Such families of applications, and the components needed to support them, form complex (sub-)ecosystems themselves; this is a common feature in technology~\cite[Ch. ``Structural Deepening'']{design:book/Arthur09}. 
To exemplify the big data ecosystem focusing on MapReduce, the figure emphasizes components in the bottom three layers, which are typically not under the control of the application developer but must nevertheless perform well to offer good non-functional properties, including performance, scalability, and reliability. 
This is due to {\it vicissitude}~\cite{DBLP:conf/ccgrid/GhitCHHEI14} in processing data in such ecosystem, that is, the presence of workflows of tasks that are arbitrarily compute- and data-intensive, and of unseen dependencies and (non-)functional issues.

{\bf Examples of large-scale computer ecosystems:}
Unlike their constituents, ecosystems are difficult to identify precisely, because their limits have not been defined at design time, or shared in a single software repository or hardware blueprint.
Large-scale examples of computer ecosystems include: (i) the over 1,000 Apache cloud and big data components published as open-source Apache-licensed software\footnote{\url{https://github.com/apache}}, (ii) the Amazon AWS cloud ecosystem, which is further populated by companies running exclusively on the AWS infrastructure, such as Netflix\footnote{\url{https://github.com/netflix}}, (iii) the emerging ecosystem built around the MapReduce and Spark\footnote{\url{https://github.com/databricks}} big-data processing systems.

{\bf When is a system not an ecosystem?}
Under our definition, not every system can be an ecosystem, and even some advanced systems do not qualify as ecosystems, including:
(i) existing {\it audited} systems are rarely built as ecosystems, and especially avoid including multi-party software and too autonomous components, (ii) legacy monolithic systems with tightly coupled components, (iii) legacy systems developed with relatively modern software engineering practices, but which do not consider the sophisticated non-functional requirements of modern stakeholders, (iv) systems developed for a specific customer or a specific  business unit of an organization, which now need to offer open-access for many and diverse clients.

It may not be possible to distinguish for all existing systems whether they are within the scope of this work's definition of ecosystems. 
This type of ambiguity exists in the definition of many new domains of computer science that are not tightly coupled to a specific technology, including embedded systems, meta-computing and grid computing systems, cloud computing systems, and big data systems. The ambiguity allows these fields to be diverse and useful, as rich field of science and engineering.

\subsection{Fundamental Problems of Ecosystems}
\label{sec:bg:problems}

{\it The first fundamental problem} is that we lack the systematic laws and theories to explain and predict the large-scale, complex operation and evolution of computer ecosystems. For example, when an ecosystem under-performs or fails to meet increasingly more sophisticated non-functional requirements, customers stop using the service~\cite{DBLP:journals/behaviourIT/Nah04,taylor2005death}, but currently we do not have the models to predict such under-performing situations, or the instruments to infer what could happen, even for simple ecosystems comprised of small combinations of arbitrary distributed systems.

{\it The second fundamental problem} is that we lack the comprehensive technology 
to maintain the current computer ecosystems.
For example, we know from grid computing the damage that a failure can trigger in the entire computer ecosystem~\cite{DBLP:conf/grid/IosupJSE07,DBLP:conf/europar/GalletYJKIE10,DBLP:conf/grid/YigitbasiGKIE10}, and far all the large cloud operators, including Amazon, Alibaba, Google, Microsoft, etc., 
have suffered significant outages~\cite{DBLP:conf/cloud/GunawiHSLSAE16} and SLA issues~\cite{taylor2005death} despite extensive site reliability teams and considerable intellectual abilities. In turn, these outages have correlated failures, as for example experienced when drafting this and other articles on the Amazon-based Overleaf.
Moreover, we seem to have opened a Pandora's Box of poorly designed systems, which turned into targets and sources of cyberattacks (e.g., 
hacking\footnote{Examples: \url{https://www.washingtonpost.com/world/national-security/israel-hacked-kaspersky-then-tipped-the-nsa-that-its-tools-had-been-breached/2017/10/10/d48ce774-aa95-11e7-850e-2bdd1236be5d_story.html}, \url{http://www.bbc.com/news/technology-42056555}, \url{https://www.volkskrant.nl/tech/dutch-agencies-provide-crucial-intel-about-russia-s-interference-in-us-elections~a4561913/}}, 
ransomware\footnote{Examples: \url{https://arstechnica.com/information-technology/2017/05/virulent-wcry-ransomware-worm-may-have-north-koreas-fingerprints-on-it/}, \url{https://www.theverge.com/2016/11/27/13758412/hackers-san-francisco-light-rail-system-ransomware-cybersecurity-muni}}, 
malware~\cite{DBLP:conf/icdcs/YangZG17}, and botnets~\cite{DBLP:conf/uss/GuPZL08}).

{\it The third fundamental problem} is that we are not equipped to explore the future of computer ecosystems, and in particular we cannot now design, tune, and operate the computer ecosystems that can seamlessly support all the societally relevant application domains, to the point where 
multiple, possibly competitive, organizations and individuals can use computing as an utility (similarly to the electricity grid, including its local shifts toward decentralized smart grids) or as a service (as for the logistics and transportation industry). 
For example, sophisticated users are already demanding but not receiving detailed control over heterogeneous resources and services, the right to co-design services with functional requirements offered through everything as a service~\cite{DBLP:conf/IEEEcloud/DuanFZSNH15}, and the opportunity to control detailed facets of non-functional characteristics such as risk management, performance isolation, and elasticity~\cite{DBLP:journals/corr/HerbstKOKEIK16}.

{\it The fourth fundamental problem} is that of 
peopleware, 
especially because 
the personnel designing, developing, and operating these computer ecosystems already numbers millions of people world-wide but is severely understaffed and with insufficient replacement available~\cite{tr:ec14cloud}.

{\it The fifth fundamental problem} is participating in the emerging practice beyond mere technology. 
Compounding the other problems, the \DistribSys{} community seems to focus excessively on technology, a separation of concerns that was perhaps justifiable but is becoming self-defeating. This focus has brought until now important advantages in producing rapidly many successful ecosystems, but is starting to have important drawbacks: 
(i) we have to answer difficult, interdisciplinary questions about how our systems influence the modern society and its most vulnerable individuals~\cite{issues:Chakrabarti18}, and in general about the emergence of human factors such as (anti-)social behavior~\cite{DBLP:conf/netgames/MartensSIK15},
(ii) we have to investigate general and specific questions about the evolution of systems, including how the knowledge and skill have concentrated in relatively few large-scale ecosystems?, and what to do, and with which interdisciplinary toolkit, to prevent this from hurting competition and future innovation~\cite{issues:Economist18a}?

%% file: content-main-core.tex
\section{\ourdomain{} (\ourdomainshort)}
\label{sec:core}

In this section, we introduce the fundamental concepts and principles of \ourdomainshort{}.
We explain its background, give a definition of \ourdomainshort{}, explain its goal and central premise, and focus on key aspects of this domain. We explain how \ourdomainshort{} extends the focus of traditional \DistribSys{}, and how it synthesizes research methods from other related domains.

\subsection{What Is \ourdomainshort?}
\label{sec:core:mcs}

We now define \ourdomainshort{} as a use-inspired discipline~\cite{DBLP:journals/cacm/Snir11}:
\definition{
\ourdomainshort{} focuses on the science, design, and engineering of ecosystems. It aims to understand ecosystems and to make them useful to the society. 
}

\begin{table}[!t]\centering
\ra{1.1}
\begin{tabular}{@{}lll@{}} \toprule
\multicolumn{3}{c}{\ourdomain{} ($\S$\ref{sec:core:mcs})} \\
\midrule
Who? & Stakeholders       &  scientists, engineers, designers, others \\
\midrule
What? & Central Paradigm  &  properties derived from ecosystem  \\
      & Focus             &  structure, organization, and dynamics  \\
      & Concerns          &  functional and non-functional properties  \\
      &                   &  emergence, evolution \\
\midrule
How?  & Design            &  design methods and processes  \\
      & Quantitative      & measurement, observation  \\
	  & Exper. \& Sim. & methodology, TRL, benchmarking  \\
	  & Empirical & correlation, causality iff. possible  \\	  
	  & Instrumentation   & experiment infrastructure  \\ 
	  & Formal models & validated, calibrated, robust  \\ 
\midrule
Related  & Computer science  &  Distrib.Sys., Sw.Eng., Perf.Eng.\\
($\S$\ref{sec:core:thisscience})      & Systems/complexity   &  General Systems Theory, etc.  \\
	  & Problem solving  &  computer-centric, human-centric  \\
\bottomrule
\end{tabular}
\caption{An overview of \ourdomainshort{}.}
\label{tab:domain:mcs}
\end{table}

Table~\ref{tab:domain:mcs} summarizes \ourdomainshort{}: Who? What? How? Which other core issues? (all addressed in this section) and What are the related concepts \ourdomainshort{} draws from? (addressed in Section~\ref{sec:core:thisscience}). We now elaborate on each part, in turn.

{\bf Who? Stakeholders:} \ourdomainshort{} involves a large number of stakeholders, characteristic and necessary for a domain that applies to diverse problems with numerous users. We consider explicitly the scientists, engineers, and designers of \ourdomainshort{} systems involved in solving the numerous challenges of the field (discussed in Section~\ref{sec:challenges}) and in using results in practice, the industry clients and their diverse applications (Section~\ref{sec:use}), the governance and legal stakeholders, etc. We also consider as stakeholders the population: individuals at-large, as clients and as (life-long) students.

\namedbox{Goal}{
The goal of \ourdomainshort{} is to understand and eventually control complex ecosystems and, recursively, their constituent parts, thus satisficing possibly dynamic requirements and turning ecosystems into efficient utilities. To this end, \ourdomainshort{} must explain how and why the ecosystem differs, functionally and non-functionally, from mere composition of its constituents.
}

{\bf What? The Central Premise:} 
\ourdomainshort{} starts from the premise that the interaction between systems in an ecosystem, and the way the ecosystems stakeholders interact with the ecosystem (and among themselves), drives to a large extent the operation and characteristics of the ecosystem.
Thus, 
\ourdomainshort{} focuses explicitly on the \textit{structure, organization, and dynamics} of systems when operating in assemblies, hierarchies, and larger ecosystems, rather than understanding and building single systems working in isolation. 

Both the \textit{functional and the non-functional properties} of these ecosystems, and recursively of their constituent systems, are central to understanding and engineering ecosystems.

Over periods of time both that are short (seconds to days) or long (weeks to years), 
ecosystems may experience various forms of \textit{emergent and chaotic behavior}, and of {\it evolution} (discussed in the following).
Understanding emergent and evolutionary behavior, and controlling  it subject to efficiency\footnote{Although process economics is better equipped than \ourdomainshort{} to address costing, pricing, and utility functions, in practice designers and engineers are expected to conduct or at least provide quantitative input for these tasks.} considerations, is also central to \ourdomainshort{}.

{\bf How? A general approach and methodology:} To begin work on \ourdomainshort{}, we consider the following elements that will need to be adapted, extended, and created for computer ecosystems, and ultimately will result in new approaches and methodologies:
(i) methods and processes characteristic to design~\cite{design:book/Lawson05,design:book/Cross11}, and design science applied to information systems~\cite{DBLP:journals/misq/HevnerMPR04} and to the design of (computer) systems;
(ii) quantitative research, in particular collection of data through measurement and (longitudinal) observation, statistical modeling of workloads~\cite{DBLP:journals/internet/IosupE11}, failures~\cite{DBLP:conf/grid/YigitbasiGKIE10,DBLP:conf/europar/GalletYJKIE10}, and reaching formal (analytical) models;
(iii) experimental research, including real-world experimentation through prototypes, and simulation, both under realistic
workload conditions and even under community-wide benchmarking settings;
(iv) empirical and phenomenological research, including qualitative research resulting in comprehensive surveys~\cite{journals/jss/BreretonKBTK07} and field surveys;
(v) modern system evaluation, using instrumentation beyond what is needed to test typical \DistribSys{} (e.g., large-scale infrastructure comparable with medium-scale industry infrastructure~\cite{DBLP:journals/computer/BalELNRSSW16}), focusing on an extended array of indicators and metrics (e.g., performance, availability, cost, risk, various forms of elasticity~\cite{DBLP:journals/corr/HerbstKOKEIK16}), and developing approaches for meaningful comparison across many alternatives for the same component~\cite{DBLP:journals/pvldb/IosupHNHPMCCSAT16} or policy point~\cite{journal/tompecs/IlyushkinAHPEI18}.

{\bf How? Other issues:}
We envision several other core issues important for \ourdomainshort{}:
(i) peopleware: processes for training, educating, engaging people, especially the next generation of scientists, designers, and engineers,
(ii) making available free and open-access artifacts, both open-source software and common-format data, 
(iii) ensuring a balance of recognition between scientific, design, and engineering outcomes, across the community,
and (iv) ethics and other interdisciplinary issues.

\subsection{More on the Central Premise}
\label{sec:core:mcs:centralpremise}

Among the core aspects of the central premise, we see the structure, organization, and dynamics of ecosystems, and the functional and non-functional properties as being derived and expanded directly from \DistribSys{} community, with the main difference being that we focus here on the larger, more complex ecosystems\footnote{Understanding assemblies where components are provided by different developers, and used by multiple stakeholders, is challenging.}. We now elaborate in turn on two distinguishing aspects of the central premise, emergence and chaotic behavior, and evolution. 

{\bf Emergence and chaotic behavior}, both functional\footnote{DNS tunneling~\cite{DBLP:journals/igpl/AielloMP13} is just one of the many examples of changing the function of a design: here, from facilitating access to the Web, to enabling arbitrary Internet traffic and thus significant security breaches. Because the ecosystem is already too complex to supervise, it turns out DNS tunneling is also not a prime target of automated protection.} and non-functional\footnote{For example, in the field of big data, the community is starting to understand ecosystem performance as a complex function of Varbanescu's ``P-A-D Triangle'' (i.e., platform, algorithm including data structures, and dataset). We have tested this empirically for graph processing~\cite{DBLP:conf/ipps/GuoBVIMW14,DBLP:conf/ccgrid/GuoVIE15}.}, due to humans use or other non-deterministic elements. 
Beyond classic emergence from  Complex Adaptive Systems and the related domains of General Systems Theory (see Section~\ref{sec:core:thisscience}), we consider within the scope of \ourdomainshort{} various biologically and socially inspired mechanisms of non-technical behavior that may change the needs and thus use of the system, such as exaptation~\cite{journals/paleobio/GouldV82}, social~\cite{DBLP:journals/tkdd/JiaSBIKE15} and meta~\cite{conf/MMVE/ShenI11,DBLP:journals/tomccap/JiaSEI16} use of systems, toxicity~\cite{DBLP:conf/netgames/MartensSIK15} and other disruptive behavior, etc. 

{\bf Evolution:}
Over long periods time, \ourdomainshort{} ecosystems evolve through internal (technology push) and external (society pull) pressures. The mechanisms of evolution include~\cite[Ch.9]{design:book/Arthur09}: combining components into larger assemblies, removing redundant or useless components, replacing components with more advanced components, bridging between components and adapting the end-points of components, adding new components to address new functions and new non-functional requirements, etc. Importantly, like Arthur we envision that {\it ecosystem evolution can be at times Darwinian}, that is, incremental, selecting and varying closely related components of pre-existing technology, with the better approaches propagating over technology generations; 
but also that {\it ecosystem evolution can be non-Darwinian}, that is, radically different and abrupt, combining seemingly unrelated technology and/or addressing novel needs, with seemingly random events---which ecosystem adopted the technology first, which individual co-sponsored the invention, how quickly it started to gain market share and other soft lock-in elements---contributing to the propagation of the technology. 
The mechanisms of ecosystem evolution are within the scope of \ourdomainshort{}.

\subsection{More on the General Approach}
\label{sec:core:mcs:approach}

{\bf Design:} 
By definition, \ourdomainshort{} employs a diverse body of knowledge and skill typical to modern {\it science} and {\it engineering}, from which we further distinguish {\it design}\footnote{ We adopt here the argument made by Cross in the 1970s, and extended by Lawson~\cite[Ch.8, loc.2414, and Ch.16, loc.4988]{design:book/Lawson05}, that design is a distinct way of thinking about real-world problems with high degree of uncertainty, and of solving them: problems and solutions {\it co-evolve}.}.
The work we conduct in this field  
aims to go beyond random walks, and direct application or replication of prior work, 
We aim to establish {\it design methods and processes}, based on principles and on instruments, that meet the goal of \ourdomainshort{}.
We envision here, as a first step, adapting and extending techniques from the design of information systems~\cite{DBLP:journals/misq/HevnerMPR04} and of computer systems, and also from design not related to computers~\cite{design:book/Cross11,engineering:book/Vincenti90} or even to technology~\cite{design:book/Lawson05}.

{\bf Quantitative results:} Obtain quantitative, predictive, actionable understanding about the sophisticated functional and non-functional properties of ecosystems, and about their dynamics. It is here that advances in \PerfEng, especially {\it measurement} and statistically sound {\it observation}, can help the domain of \ourdomainshort{} get started. Specifically, collecting data from running ecosystems and from experimental settings, both real-world and simulated (see following heading), we can start accumulating {\it knowledge}. (The step to {\it understanding} cannot be fully automated, because it is dependent on the imagination of the people in the loop.) These would lead to observational models, and, later, possibly also to calibrated mechanistic models and full-system (weakly emergent~\cite[p.171]{book/Simon96}) models.

{\bf Experimentation and simulation:}
\ourdomainshort{} depends on methodologically sound real-world\footnote{\ourdomainshort{} follows the multi-decade tradition of experimental computer science~\cite{DBLP:journals/csur/Plaice95,DBLP:journals/computer/Tichy98,experimental:Feitelson06} and \DistribSys{}~\cite{experimental:Epema11,DBLP:journals/computer/BalELNRSSW16}.} and simulation-based\footnote{Simon makes a compelling case that simulation can lead to new understanding, of both computer systems about which we know much and about which we do not~\cite[Section ``Understanding by Simulating'']{book/Simon96}. He refutes that a simulator is ``no better than the assumptions built into it'', that they cannot reveal unexpected aspects, that they only apply for systems whose laws of operation we already know.} experiments, which have complementary strengths and weaknesses but combined can provide essential feedback to scientists, engineers, and designers.
Experimentation is valuable in validating and demonstrating the {\it technology-readiness level}~(TRL)\footnote{\url{http://www.earto.eu/fileadmin/content/03_Publications/The_TRL_Scale_as_a_R_I_Policy_Tool_-_EARTO_Recommendations_-_Final.pdf}} of various concepts and theories, using {\it prototypes} or even higher-TRL artifacts running preferably in real-world environments\footnote{Like Tichy~\cite{DBLP:journals/computer/Tichy98}, we disagree that mere demonstrations and proof-of-concepts can replace experimentation and simulation, even if they prove valuable for engineering products and educating stakeholders.}, in providing calibration and measurement data, in revealing aspects that we have not considered before, etc. {\it Benchmarking}, a subfield of experimentation, focuses the community on a set of common processes, knowledge, and instrumentation. Good benchmarks often make experimentation also more affordable and fair, by establishing for the community a set of meaningful yet tractable experiments.
Simulation is useful in investigating and comparing known and new designs, and dynamics including non-deterministic behavior, over long periods of simulated-time. Simulation, and to some extent also real-world experimentation, can also be used to replay interesting conditions from the past, giving the human in the loop more time and more instruments to understand.

{\bf Empirical (correlation), and if possible also phenomenological (causal), research} is necessary\footnote{As a matter of pragmatism, our empirical research may need to be data-driven (that is, {\it discovery science}~\cite{biosys:IdekerGH01}), instead of hypothesis-driven, simply because the complexity of the problems seems to exceed the capabilities of the unaided human mind. This is also the case made since the mid-2000s by \BioSys{}~\cite{biosys:IdekerGH01,biosys:Kitano02}\cite[Ch.1]{biosys:book/AlberghinaW05} and other sciences.}, if we are to understand and control especially the emergent properties of ecosystems.
Observation and measurement, and experimentation and simulation of ecosystems already are empirical methods, with their benefits and drawbacks, for studying and engineering the systems comprising the ecosystems. 
Additionally, \ourdomainshort{} must also study empirically the highly variable, possibly non-deterministic processes that include humans: 
their use of ecosystems and their new (practical) problems with using ecosystems, and 
their study, design, and engineering of ecosystems. This latter part is much less developed in \DistribSys{}, but a rise in empirical methods in \SwEng{}~\cite{DBLP:journals/computer/Tichy98,empirical:Meyer18}
and in design sciences~\cite[Ch.1]{design:book/Cross11} already employs:
studying the artifacts themselves (e.g., with static code analysis),
interviews with designers, 
observations and case studies of one or several design projects, experimental studies typically of synthetic projects, 
simulation by letting computers try to design and observing the results, and 
reflecting and thinking about own experience. 
The benefits of using these methods include deeper, including practical, understanding. The dangers include relying on ``soft methods''~\cite{DBLP:journals/computer/Tichy98} and ignoring the ``threats to validity''~\cite{empirical:Meyer18}.

{\bf Instrumentation:}
Similarly to other Big Science domains, such as astrophysics, high-energy physics, genomics and systems biology, and many other domains reliant today on e-Science, \ourdomainshort{} requires significant instrumentation. 
It needs adequate environments to experiment in, for example, the DAS-5 in the Netherlands~\cite{DBLP:journals/computer/BalELNRSSW16} and Grid'5000 in France. As in the other natural sciences, creating these instruments can lead to numerous advances in science and engineering; moreover, these instruments are ecosystems themselves and thus an {\it endogenous} object of study for \ourdomainshort.
\ourdomainshort{} also needs the infrastructure needed to complement the human mind in the task of understanding the data collected about ecosystems, to generate hypotheses automatically, and to preserve this data for future generations of scientists, designers, and engineers.

{\bf Formal (analytical) models:}
We envision that a complex set of formal mathematical models, validated and calibrated with long-term data, robust and with explanatory power beyond past data, will emerge over time to support \ourdomainshort{}. Such models will likely be hierarchical, componentized. 
The key challenge to overcome for meaningful, predictive modeling is to support the dynamic, deeply hierarchical, emergent nature of modern ecosystems. 
There may not be a steady-state, for example when users seem to behave chaotically, or high resource utilization triggers bursty resource (re-)leases in clouds.

{\it Models at different levels} must support ordinary and partial differential equations (ODEs and PDEs have multiple independent control-variables), time-dependent evolution and events, discrete states and Boolean logic, stochastic properties for each component and behavior, and capture emergent and feedback-based behavior (collectively, forms of {\it ecosystem-wide non-linearity}).
Unlike other models used in traditional and computational sciences, models in \ourdomainshort{} will also need to capture the human-created design principles and processes underlying the ecosystems, including their non-Darwinian evolution~\cite[Ch.6, loc.1875]{design:book/Arthur09}.
Thus, the emerging models will likely be complex, unlike the first-order approximations of classical physics, and may require computers to manipulate. Even then, the curse of dimensionality, i.e., too many states and parameters to explore, may make these models intractable for online predictions.

\subsection{More on Other Issues}
\label{sec:core:mcs:other}

The \DistribSys{} community seems to have already agreed on new education processes and is making progress toward our notions of peopleware support. It has also agreed that the release of software~\cite{DBLP:journals/dagstuhl-manifestos/AllenABCCCCCGGG17} and data artifacts is beneficial, although the funding and recognition are still lagging behind. \ourdomainshort{} can build on this agreement and focus in this context on computer ecosystems. We now focus on the third and fourth issues.

{\bf The balance of recognition:}
It is now common in the computing community, but hurtful to both the results and to the community itself, to consider science above engineering\footnote{It was and seems to remain common for science to dismiss engineering as merely an applied science, in general~\cite{engineering:Boon06}~\cite[p.3-4]{engineering:book/Vincenti90}.} or vice-versa\footnote{This appears to be a reverse process, in which engineers see scientific theories as overly idealistic, abstract, and ignorant of actual conditions~\cite{engineering:Boon06}. Anecdotally, Andy Tanenbaum, then a student close to the early development of the time-sharing systems at MIT, recounts that the systems community of the time had little to do with the contemporary theoretical advances in queueing theory and modeling. Later, when starting Minix, he was leading a team trying to make a running and useful distributed computer, rather than respond to needs arising from the scientific community~\cite{DBLP:journals/cacm/Tanenbaum16}. (Also personal communication, March 2017.)}, or to dismiss that design can be an independent task\footnote{Defining design as an engineering task is countered by~\cite{design:book/Cross11} and~\cite{book/Monteiro12}.}. In contrast, \ourdomainshort{} explicitly postulates that all jobs in this domain resulting in meaningful knowledge and technology are equally inspiring and useful, and thus should be equally prestigious. 
Science in this domain discovers artificial phenomena to be used in ecosystems, and thus operates in the continuum between curiosity-driven and applied, and is most commonly {\it use-inspired} in the sense of Pasteur Quadrant in the context of computer science, as analyzed by Snir~\cite{DBLP:journals/cacm/Snir11}. 
Engineering in this domain is not mere application of recipes; it requires considerable creativity, skill, and knowledge beyond traditionally scientific. 
The third component at the core of \ourdomainshort{}, (concept) design, deserves the awe inspired in our society by the creative arts, and the respect deserved for solving complex problems.

{\bf Ethics and other interdisciplinary issues:} Beyond the balance of recognition, we see many issues where historical aspects and ethics influence the evolution and development of computer ecosystems. We envision here an interdisciplinary community that engages \ourdomainshort{} practitioners, including \DistribSys{} experts, about the principles and the technology of computer systems; we see here as useful the multidisciplinary invitation sent by the Dagstuhl Seminar on the History of Software Engineering~(1996)~[p.1]\cite{DBLP:journals/dagstuhl-reports/AsprayKP96}.

\begin{figure*}[!tbh]
\centering
	\includegraphics[width=0.6\textwidth]{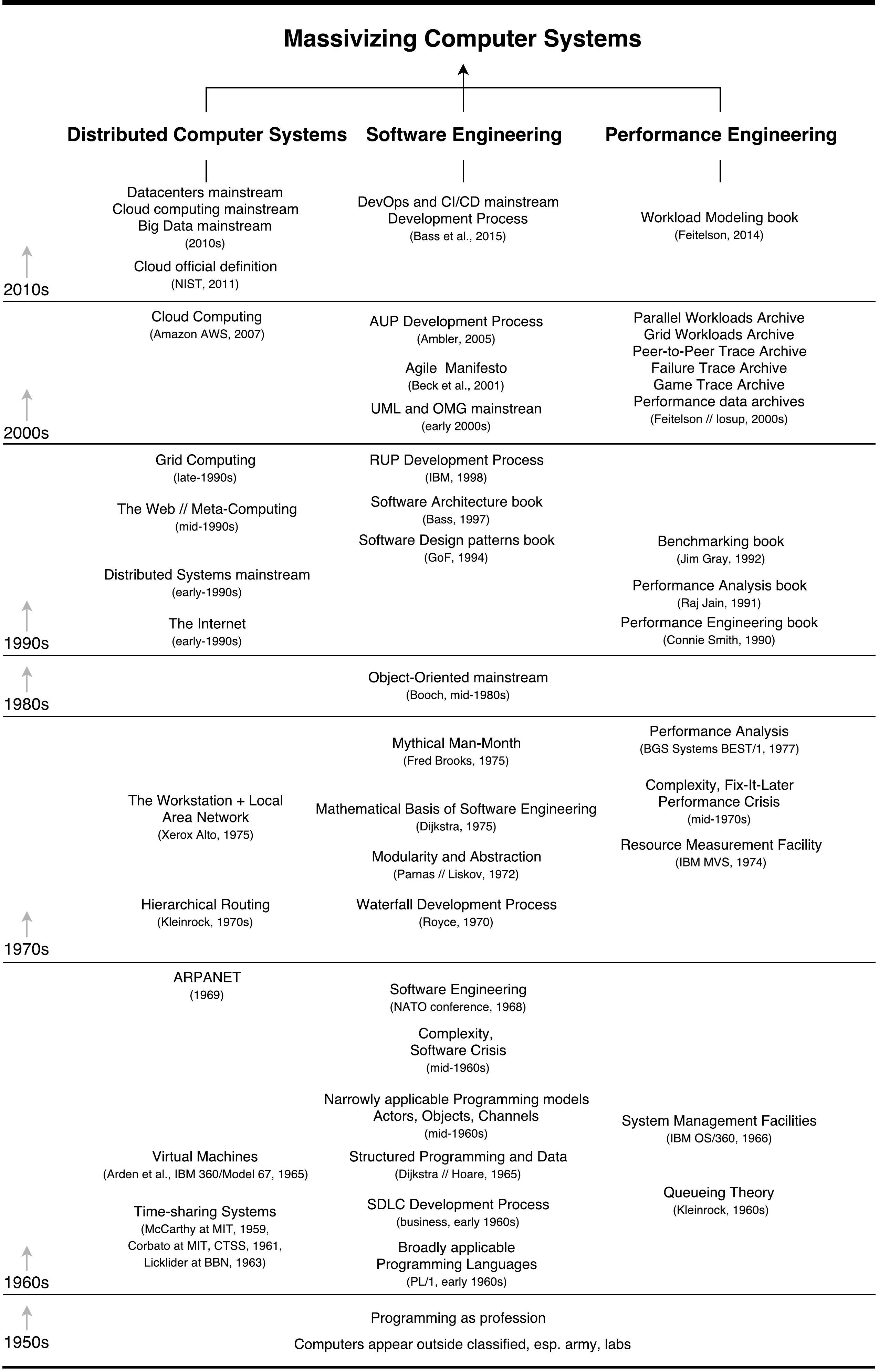}
	\caption{Main technologies leading to \ourdomainshort{}. \ourdomainshort{} is a response to the {\it ecosystems crisis} of late-2010s (see Section~\ref{sec:intro}).}
	\label{fig:history}
\end{figure*}

\subsection{How Far Are We Already?}
\label{sec:core:thisscience}

To understand the extent of progress we have made in \ourdomainshort{}, we need to understand both what techniques and processes the field is comprised of already (discussed in this section), and what applications it can have (in Section~\ref{sec:use}). Overall, \ourdomainshort{} has a large, valuable body of knowledge to build upon, which brings both the opportunity of having a diverse, tested toolbox and the complex challenge of learning and using it. Figure~\ref{fig:history} depicts the evolution in this sense of technology, in \DistribSys{} and in the complementary fields of \SwEng{} and \PerfEng{}. 

{\bf Common fields of computer science:}
We see \ourdomain{} as derived from Distributed Systems, which in turn are derived from core Computer Systems. Additionally, \ourdomain{} aims to synthesize interdisciplinary knowledge and skills primarily from Software Engineering and Performance Engineering. This is in agreement with Snir's view that computer science is ``one broad discipline, with strong interactions between its various components''~\cite{DBLP:journals/cacm/Snir11}, under which subdisciplines reinforce each other, and multidisciplinary and interdisciplinary research and practice further enable the profession. 

We have compiled a non-exhaustive list of principles and concepts \ourdomainshort{} can import from established domains: (i) from \DistribSys{}, scalability as a grand challenge extended to the concept of elasticity, communication as first-class concern, resource management including migration of workload and sharing of resources, scheduling policies and routing disciplines especially full automation, computational models including CSP and Valiant's BSP, geo-distribution especially through replication and sharding, the CAP theorem with related theoretical and practical work, concurrency, etc.;
(ii) from \CompSys{}: hierarchy as basic architecture, the modularity principle, the locality principle, the principle of separation mechanism-policy, the separation of data and process, core workload models such as workflows and dataflows, plus basic AI and machine-learning techniques used for feedback and control loops (e.g., pattern recognition, signal classification, deep learning and CNNs, Bayesian inference, expert systems), etc.; 
(iii) from \SwEng{}: data structures, algorithms, code and architectural patterns for software, processes for software engineering including testing, etc.;
(iv) from \PerfEng{}: many empirical processes, the concept of non-functional properties as first-class concern, and instruments and tools to monitor, measure, analyze, model, and predict performance, etc. 

{\bf Generalized systems and complexity theory:}
We consider as important especially for the theoretical development of \ourdomainshort{} the concepts and techniques from Complex Adaptive Systems, and the related domains of General Systems Theory, Chaos Theory, Catastrophe Theory, Hierarchical Theory, etc.: networks, non-linear effects, non-stationary processes, control, etc. However, we are also aware that much distance must be covered between theory and practice, related to these fields.

{\bf Generalized problem-solving:}
For theories and techniques of problem solving and problem satisficing\footnote{Satisficing~\cite[p.28]{book/Simon96} is about finding a solution that meets a set of requirements based on a threshold (``better than X''), instead of the goal of optimization to find an optimum (``the absolute best'').}, we consider two classes of techniques: {\it computer-centric} and {\it human-centric}.

For the former, we identify two wide-ranging and thoroughly investigated approaches: satisficing using heuristics, and solving or optimizing for simplified models. Approximate solutions generated via heuristics are generally preferred when finding optimal solutions is considered intractable\footnote{For example, when the time to solution is superpolynomial.}. 

Possibly the most widely used family of methods to investigate large solution spaces are the \emph{A*} algorithm and its optimizations, such as the \emph{iterative deepening A*}. Such methods have been refined by the \emph{artificial intelligence} community by using guided and procedural search, and developed into new fields of study, such as evolutionary computing~\cite{eiben2003introduction}, which describes a wide variety of biology-inspired search algorithms: genetic algorithms, genetic programming, particle-swarm optimization, learning classifier systems, etc. In domains where data is abundant, data mining and machine learning techniques~\cite{valiant2013probably} leverage good results by extracting knowledge or building predictive models from the available data. Simple heuristics addressing highly specialized problems appear in control theory, with practical applications for relatively simple mechanical systems.

In domains where simplified (mathematical) models can be drawn, finding (near-)optimal solutions becomes less difficult than "blindly" exploring large search spaces. The simpler and most widely used models is the basic linear (integer) programming method, or the dynamic programming paradigm used for finding (near-)optimal solutions when the solution space can be bounded and well-defined. This set of simpler models also includes rule-based expert systems, where a knowledge base is used as inferencing engine. More complex models, as the ones defined by queuing theory led to seminal results such as \emph{Little's Law}, widely used in distributed systems, networking and scheduling. Models have also been used successfully for performance analysis and prediction. Frameworks such as the Roofline model~\cite{DBLP:journals/cacm/WilliamsWP09} are effective in predicting the performance achieved by modern multicore architectures using only modest numbers of parameters (e.g., memory bandwidth, floating-poing performance, operational intensity).  

{\it Human-centric}: Because many of the \ourdomainshort{} still need design and tuning, and because in deployed \ourdomainshort{} systems it is common to have humans-in-the-loop, we also consider and plan on educating people about human-centric approaches for problem solving as applied in \ourdomainshort{}. Combining the taxonomies proposed by Beitz et al.~\cite{design:book/BeitzPFG07} and by Shah et al.~\cite{journals/designstudies/ShahSV03}, we consider {\it intuitive} and {\it discursive} (that is, recipe-based) techniques. Among the intuitive techniques are: the brainstorming of Osborn, the gallery method of Hellfritz, the lateral-thinking method of DeBono, storyboarding, Fishborne/Ishikawa cause-and-effect diagrams, the synectics of Gordon, etc. 
Among the discursive techniques, we consider history-based techniques such as TRIZ by Altshuller, the (general) morphological analysis of Zwicky, design catalogs and comprehensive surveys, etc.; and analytical techniques such as selection and evaluation using systematic charts, single- and multi-values of merit for rating (and benchmarking) systems, use-value analysis and utility functions, pair-wise tournaments and competitions, etc.

\section{Ten Core Principles of \ourdomainshort}
\label{sec:core:principles}

We introduce in this section ten core principles of \ourdomainshort{}. Our principles are not focusing on the details of building a particular system or ecosystem. Instead, they focus on understanding the higher principles that can shape how the computer ecosystems we envision are related to a science of systems, peopleware, and methodology (meta-science) enabling them.
Any attempt to formulate a fixed number of principles is artificial, but it can help guide the development of a scientific domain or field of practice\footnote{As did the Agile Manifesto's 12 principles (\url{agilemanifesto.org}).}.

\begin{table}[!t]\centering
\ra{1.1}
\begin{tabular}{@{}lrl@{}} \toprule
& \multicolumn{2}{l}{Principle} \\ \cmidrule(r){2-3}
Type & Index & Key aspects \\ 
\midrule
Systems ($\S$\ref{sec:core:system}) & P\ref{pr:ecosystems}&  The Age of Ecosystems \\
 & P\ref{pr:software}      &  software-defined everything\\
 & P\ref{pr:nfr}           &  non-functional requirements\\
 & P\ref{pr:rms}           &  RM\&S, Self-Awareness\\
 & P\ref{pr:artcrft}       &  super-distributed\\
\midrule
Peopleware ($\S$\ref{sec:core:peopleware}) & P\ref{pr:educating}&  fundamental rights \\
 & P\ref{pr:personalprivilege} &  professional privilege\\
\midrule
Methodology ($\S$\ref{sec:core:methodology}) & P\ref{pr:reproducible}&  science, practice, and culture of \ourdomainshort{}\\
 & P\ref{pr:evolution}     &  evolution and emergence\\
 & P\ref{pr:ethics}        &  ethics and transparency\\
\bottomrule
\end{tabular}
\caption{The 10 key principles of \ourdomainshort{}. (Acronyms: RM\&S stands for Resource Management and Scheduling.)}
\label{tab:principles}
\end{table}

We hold as our highest principle that: 
\principle{pr:ecosystems}{{\bf This is the Age of Computer Ecosystems.}}
As indicated in Section~\ref{sec:core:ecosystems} that large-scale ecosystems are now at the core of many if not most private and public utilities; this is the Age of Computer Ecosystems. 
Derived from its goal and as stated in its central premise (see Section~\ref{sec:core:mcs}), \ourdomainshort{} aims to understand and design computer ecosystems, working efficiently at any scale, to benefit the society.
This requires a science of pragmatic, predictable, accountable computer systems that can be composed in nearly infinite ways, be controlled and understood despite the presence of complexity,  emergence, and evolution, and whose core operative skills can be taught to all people. Overall, this leads to the principles summarized by Table~\ref{tab:principles}.

\subsection{Systems Principles}
\label{sec:core:system}
\label{sec:core:organization}
\label{sec:core:characteristics}

\ourdomainshort{} proposes a non-exhaustive set of principles guiding work on computer systems and ecosystems.

\principle{pr:software}{{\bf Software-defined everything, but humans can still shape and control the loop.}}
The ecosystem is comprised of software and software-defined (virtual) hardware, which allow for advanced control capabilities and for extreme flexibility. ``Software is eating the world''\footnote{\url{https://tinyurl.com/Andreesen11}}, but under control.

However autonomous these ecosystems can become, humans must still be able to control them\footnote{Starting with the 1960s, (dystopian) sci-fi has imagined many scenarios regarding loss of privacy, of sovereignty, and ultimately of free will. Many of these are only now emerging as real-world problems.}. Techniques for ensuring human control work in parallel with increasing and even full automation, where humans delegate specific decisions for a while.
Because humans must still be in control, \ourdomainshort{} must go deeper than just building technology.

\principle{pr:nfr}{{\bf Non-functional properties are first-class concerns, composable and portable, whose  relative importance and target values are dynamic.}}
Non-functional requirements, including security, trust, privacy, scalability, elasticity, availability, performance, are first-class concerns, but {\it the importance and the characteristics of each requirement may be fluid over time, and depends on stakeholders, clients, and applications}. 

We envision guarantees of both functional and non-functional properties, however and whenever assemblies are composed, even when complexity, emergence, and evolution exist. Long-term,  
after the maturation of \ourdomainshort{}, we envision that even operational guarantees, including limits of emergence, can be ensured through the composability and portability of non-functional properties of ecosystem and system components.

Among the guarantees, we envision not only specialized service objectives/targets (SLOs) and overall agreements (SLAs), 
but also general, ecosystem-wide guarantees such as 
performance isolation (vs. performance variability), 
tolerance to vicissitude (such as workload and requirement mixes and changes), 
tolerance to correlated failures, 
tolerance to intrusion and to other security attacks, 
etc.

\principle{pr:rms}{{\bf Resource Management and Scheduling, and their combination with other capabilities to achieve local and global Self-Awareness, are key to ensure non-functional properties at runtime.}}
Resource Management and Scheduling is a key building block without which \ourdomainshort{} is not sustainable or often even  achievable. Consequently also of the scale and complexity of modern ecosystems, {\it disaggregation} and {\it re-aggregation of software and software-defined hardware} become key operations. 

{\it Self-awareness} is a key building block, without which scalability and efficiency, and many other non-functional properties, are not attainable and controllable in the long run.
Self-awareness includes monitoring and sensing, which give input (feedback) to Resource Management and Scheduling and thus lead to better (albeit slower, and possibly uncontrolled) decisions.

\principle{pr:artcrft}{{\bf Ecosystems are super-distributed.}}
Everything in \ourdomainshort{} is distributed\footnote{The list of principles of computing proposed by Denning and Martell~\cite{principles:book/DenningM15} curiously omits distribution, although it does include networking and parallelism.}. Although some ecosystems operate primarily under one human-control unit, e.g., the management of Amazon controls the Amazon AWS operations and thus also the infrastructure, these ecosystems are still comprised of a set of systems that operate under the central paradigm of \DistribSys{}: ``a collection of autonomous computing elements that appears to its users as a single coherent system''~\cite{DBLP:journals/computing/SteenT16}. 

\ourdomainshort{} ecosysems are {\it super-distributed}: 
Following our definition in Section~\ref{sec:core:ecosystems} and as with any technology~\cite{design:book/Arthur09}, ecosystems in \ourdomainshort{} are recursively distributed. This is a form of {\it super-distribution}: distributed ecosystems comprised of distributed ecosystems, in turn comprised of distributed ecosystems, etc.

Beyond the traditional concerns of \DistribSys{}, super-distribution is also concerned with many desirable {\it super-}properties: super-flexibility and super-scalability (discussed in the following), 
multiple ownership of components and federation, multi-tenancy, disaggregation and re-aggregation of systems and workloads, interoperability including the grafting of third-party systems into the ecosystem, etc. 

Extending a term from management theory~\cite[Ch.2]{concepts:book/BahramiE05}, we define {\it super-flexibility} as the ability of an ecosystem to ensure {\it both} the functional and non-functional properties associated with stability and closed systems (e.g., correctness, high performance, scalability, reliability, and security), {\it and} those associated with dynamic and open systems (e.g., elasticity, streaming and event-driven, composability and portability). Super-flexibility also introduces a framework for managing product mergers and break-ups (e.g., due to technical reasons, but also due to legal reasons such as anti-monopoly/anti-trust law) on short-notice and quickly.

Similarly to super-flexibility, {\it super-scalability} combines the properties of closed systems (e.g., weak and strong scalability) and of open systems (e.g., the many faces of elasticity~\cite{DBLP:journals/corr/HerbstKOKEIK16}). Inspired by Gray~\cite{DBLP:journals/jacm/Gray03}, we see this new form of scalability as a grand challenge in computer science.

\subsection{Peopleware Principles}
\label{sec:core:peopleware}

\ourdomainshort{} provides services to hundreds of millions of people, through ecosystems created by a large number of amateurs and professionals. 
Inspired by the software industry's struggle to manage and develop its human resources, we explicitly set principles about peopleware.

\principle{pr:educating}{{\bf People have a fundamental right to learn and to use ICT, and to understand their own use.}}
\ourdomainshort{} must lead to teachable technology: in our vision, all stakeholders of all public computer ecosystems can be taught basic ecosystems-related skills. 
For example, individuals should be able to reading their own consumption meters and understand the reading, much as they do for their other utilities such as electricity and running water.

As a warning anecdote\footnote{Kindly proposed by Dick Epema.}, the Dutch Government has tried to introduce in the past decade various broad technologies for governance, such as digital ids, digital documents, and digital voting. An important issue has proven so far the technical level required by the proposed solutions, which currently seems to exclude millions of people, especially 
old people 
and a part of the younger generation 
especially from poor and immigrant origins. 
It remains unacceptable to exclude large parts of a population from basic societal and governance services.

\principle{pr:personalprivilege}{{\bf Experimenting, creating, and operating ecosystems are professional privileges, granted through provable professional competence and integrity.}} 

To limit damage to the society and to the profession itself, everyone who experiments with, creates, or operates ecosystems that others rely on must be subject to professional checks and balances. As a community, 
we are no longer in position to argue technology in general, and especially ecosystems reaching many people, is only beneficial and thus creating and operating such technology should be done without restriction. Vardi observes ``I realized recently that computing is not a game--it is real--and it brings with it not only societal benefits, but also significant societal costs''~\cite{DBLP:journals/cacm/Vardi18}. 
This puts our field in line with medical and legal professions, but with the added pressure resulting from the increase of contract work in our field~\cite{profession:contract18}\footnote{Besides the possible increase in quacks and shams among the practitioners of our field, due to lack of verifiable credentials, contract jobs currently have lower job benefits and insurance~\cite{profession:contract18}, which can lead to pressure to accept unprofessional and even unethical requirements.}.

As has been argued about the profession of software engineering\footnote{\url{http://repository.cmu.edu/cgi/viewcontent.cgi?article=1192&context=sei}}, and later about the profession of computing in general~\cite{DBLP:journals/cacm/DenningF11} (whose terminology we follow), we need to establish a profession of Massivizing Computer Systems. This requires establishing the core roles that stakeholders can play, including the services {\it professionals} can provide to {\it clients}. Clients have the right to be protected ``from [...] own ignorance by such a professional''~\cite[loc.4338-4339]{design:book/Lawson05}. The profession sanctions, through the guidelines of a professional society, the {\it body of knowledge} and the {\it skills} used in practice, and the {\it code of ethics} of the profession. Bodies of knowledge expand through organized (scientific) disciplines, whereas skills expand through the practice of organized trades. Professional (accredited) education provides training for both, and higher education also provides training into the processes of expanding both. Trained professionals are certified and accredited, and can lose their license or worse on abuse.

To train \ourdomainshort{} professionals, two elements need to be added to the general computing-core disciplines proposed by Denning and Frailey~\cite{DBLP:journals/cacm/DenningF11}: systems thinking and design thinking.

People with Systems Thinking skills can analyze computer ecosystems to find their laws and to formulate theories of operation, and can synthesize and tune computer ecosystems.

People with technology oriented Design Thinking skills can design computer ecosystems and the interfaces that enable their interoperability, recursively across the super-distributed, super-flexible framework (see Principle~\ref{pr:artcrft}).
Systems and design thinking will foster invention and creative designs, through the work of both many practitioners (e.g., engineers), and (relatively few) scientists and designers.

\subsection{Methodological Principles}
\label{sec:core:methodology}

As a field of computer systems, itself a field of computer science, \ourdomainshort{} leverages their scientific principles, including the list compiled by Denning~\cite[p.32]{DBLP:journals/cacm/Denning13a}: (i) focusing on a pervasive phenomenon, which it tries to understand, use, and control (\ourdomainshort{} focuses on computer ecosystems); (ii) spans both artificial and natural processes regarding the phenomenon (\ourdomainshort{} both designs and studies its artifacts at-large); (iii) aims to provide meaningful and non-trivial understanding of the phenomenon; (iv) aims to achieve reproducibility, and is concerned with the falsifiability of its proposed theories and models; etc. \ourdomainshort{} also includes in its methodological principles a broader principle, related to the ethics of the profession (linked also with Principle~\ref{pr:personalprivilege}).

\principle{pr:reproducible}{{\bf We understand and create together a science, practice, and culture of computer ecosystems.}}
We envision fostering a domain of \ourdomainshort{} where {\it everything we develop is tested and benchmarked, reproducibly}. Although providing a full set of principles leading to this goal goes beyond the scope of this article\footnote{This is part of our ongoing research as part of the international SPEC Research Group, through its Cloud Group.}, we see a set of desirable steps toward this end: 
(i) {\it Reproducibility as essential service to the community:} we must mature as a science and value reproducibility studies~\cite{Feitelson:2005:Experimental}, including by publishing reproducibility studies as other domains do~\cite{baker2016nature};
(ii) {\it Open-access, open-source:} both software~\cite{DBLP:journals/dagstuhl-manifestos/AllenABCCCCCGGG17} and data artifacts are shared with all stakeholders, receiving for this just reward and recognition~\cite{DBLP:journals/corr/SmithNKBGGHMMMP17}, including appropriate levels of funding;
(iii) {\it Negative results are useful:} following an increasingly visible community in \SwEng{}~\cite{DBLP:journals/concurrency/MaheshwariKOWT17}, we postulate that past failures, especially observed through experiments that falsify predicted results, must be recorded and shared, leading to future success;
(iv) {\it Neutral results are useful:} in the current approach of the science of computer systems, it seems that results are rarely worthy of publication, unless the results are strongly positive (or, rarely, strongly negative). We envision that neutral, even if previously unknown and expanding the body of knowledge on meaningful problems\footnote{For example, consider the notion of super-flexibility. Making existing ecosystems multi-dimensionally elastic is desirable and can lead to significant reduction in operational cost. However, this may lead currently to loss of performance, and many other trade-offs~\cite{DBLP:conf/sigmetrics/GhitYIE14}. Exploring the trade-off may not yet lead to new solutions, but it is valuable for the community at large. In our experience, such studies are often rejected from top conferences.}, will receive as much opportunity for publication as the other kinds of results;
(v) {\it Laws and theories of ecosystem operation are valuable:} contrasting to what we perceive as a bias toward ``working systems'', we see an increasing need for conducting empirical and other forms of research leading to laws of operation and possibly theories derived from it.

\principle{pr:evolution}{{\bf We are aware of the evolution and emergent behavior of computer ecosystems, and control and nurture them. This also requires debate and interdisciplinary expertise.}} 
Short- and long-term evolution, and short-term emergent behavior, can shape the use of current and future ecosystems. Practitioners in \ourdomainshort{} must be aware of the evolution of system properties, requirements, and stakeholders, and strive to be aware of emergent behavior. 

We must study existing principles~\cite{principles:book/DenningM15} and revisit periodically what is valuable in our and related fields. {\bf Corollary:} this principle also requires to revisit periodically the principles of \ourdomainshort{} discussed in this section.

Constantly monitoring for evolutionary and emergent behavior in ecosystems offers important opportunities and advantages. With good hindsight, it is possible to steer and nurture the evolution of the field efficiently, by first re-using as much as possible what already exists, and only then, iff. needed, developing new concepts, theories, and ultimately new systems and ecosystems.
With early identification of emergent behavior, DevOps~\cite[p.3]{concepts:book/Bass15} can first understand, then tune or even change the system, e.g., by adding new incentives and mechanisms to steer (unwanted) human behavior~\cite{DBLP:journals/internet/IosupBSJK14,DBLP:conf/netgames/MartensSIK15}.

Adhering to this principle is challenging, at least in the complexity of combining a diverse set of methodological theories and techniques. 
For example, from methodology already in use in \DistribSys{}, \SwEng{}, and \PerfEng{}, key to \ourdomainshort{} are the art and craft of the comprehensive survey,
longitudinal studies revealing long-term system operation, 
etc.
From interdisciplinary studies, key to \ourdomainshort{} are
field surveys of common practice and its evolution, 
workshops that truly engage the experts in debate~\cite{DBLP:journals/cacm/Vardi11}, and 
involvement of society at-large in discussing the ethics and practice of the field~\cite{DBLP:journals/cacm/Vardi10}.

\principle{pr:ethics}{{\bf We consider and help develop the ethics of computer ecosystems, and inform and educate all stakeholders about them.}} 
We have already indicated in Section~\ref{sec:intro} how our focus exclusively on technology exposes the community to various ethical risks. Overall, we envision for \ourdomainshort{} {\it an ethical imperative to actually solve societal problems}, which means our focus must broaden and become more interdisciplinary, and \ourdomainshort{} must {\it develop a body of ethics to complement the body of knowledge}. As a benefit of considering ethical issues, we envision new functional and non-functional requirements to be addressed by design in a new generation of \ourdomainshort{} ecosystems.

%% file: content-main-challenges.tex
\section{Twenty Research Challenges for \ourdomainshort}
\label{sec:challenges}

Although we see well the challenges raised by the proliferation of ecosystems and especially their constituents (see Sections~\ref{sec:intro} and~\ref{sec:core:mcs}), we are just beginning to understand the difficulties of working with ecosystems instead of merely systems. Known difficulties include, but are not limited to: the sheer volume, the group and hierarchical behavior under multiple ownership and multi-tenancy, the interplay and combined action of multiple adaptive technique, the super-distributed properties, and the remaining issues captured by our principles (see Section~\ref{sec:core:principles}).

\begin{table}[!t]\centering
\ra{1.1}
\begin{tabular}{@{}lrll@{}} \toprule
& \multicolumn{3}{c}{Challenge} \\ \cmidrule(r){2-4}
Type & Index & Key aspects & Princip.\\ 
\midrule
Systems & C\ref{ch:ecosystems}&  Ecosystems, overall & P\ref{pr:ecosystems}\\
($\S$\ref{sec:ch:system}) & C\ref{ch:swdefined}     &  Software-defined everything & P\ref{pr:software}\\
 & C\ref{ch:nfr}           &  Non-functional requirements & P\ref{pr:nfr}, P\ref{pr:artcrft}\\
 & C\ref{ch:heterogeneity} &  Extreme heterogeneity & P\ref{pr:rms} \\
 & C\ref{ch:sociallyaware} &  Socially aware & P\ref{pr:rms}\\
 & C\ref{ch:adaptation}    &  Adaptation, self-awareness & P\ref{pr:rms}\\
 & C\ref{ch:scheduling}    &  Scheduling, the dual problem & P\ref{pr:rms}, P\ref{pr:artcrft}\\
 & C\ref{ch:service}       &  Sophisticated services & P\ref{pr:rms} \\
 & C\ref{ch:ecosystem}     &  The Ecosystem Navigation challenge & P\ref{pr:software}--\ref{pr:artcrft}\\
& C\ref{ch:interoperation}&  Interoperability, federation, delegation & P\ref{pr:rms}, P\ref{pr:artcrft}\\

\midrule

Peopleware  & C\ref{ch:engagement}    &  Community engagement & P\ref{pr:educating}\\
($\S$\ref{sec:ch:peopleware}) & C\ref{ch:curriculum}    &  Curriculum, BOKMCS & P\ref{pr:educating}\\
 & C\ref{ch:explaining}    &  Explaining to all stakeholders & P\ref{pr:rms}, P\ref{pr:educating}\\

 & C\ref{ch:design}        &  The Design of Design challenge & P\ref{pr:educating}, P\ref{pr:personalprivilege}\\

\midrule
Methodology & C\ref{ch:reproducible}& Simulation and  & P\ref{pr:personalprivilege}, P\ref{pr:reproducible}\\
($\S$\ref{sec:ch:methodology}) & & Real-world experimentation & \\
 & C\ref{ch:reproresults}  &  Reproducibility and benchmarking & P\ref{pr:personalprivilege}, P\ref{pr:reproducible}\\
 & C\ref{ch:testing}       &  Testing, validation, verification & P\ref{pr:reproducible}\\
 & C\ref{ch:science}       &  A Science of \ourdomainshort{} & P\ref{pr:reproducible}, P\ref{pr:evolution}\\
 & C\ref{ch:newworld}      &  The New World challenge & P\ref{pr:reproducible}, P\ref{pr:evolution}\\
 & C\ref{ch:ethics}        &  The ethics of \ourdomainshort{} & P\ref{pr:ethics}\\
\bottomrule
\end{tabular}
\caption{A shortlist of the challenges raised by \ourdomainshort{}.}
\label{tab:challenges}
\end{table}

\challenge{ch:ecosystems}{\refpr{pr:ecosystems}}{Ecosystems instead of systems.}

We see as the grand challenge of \ourdomainshort{} re-focusing on entire ecosystems:

How to take ecosystem-wide views? How to understand, design, implement, deploy, and operate ecosystems? How to balance so many needs and capabilities? How to support so many types of stakeholders? How do the challenges raised by ecosystems co-evolve with their solutions? What new properties will emerge in ecosystems at-large and how to address them? 
These and similar questions raise numerous challenges related to systems~(see Section~\ref{sec:ch:system}), peopleware~(see Section~\ref{sec:ch:peopleware}), and methodology~(see Section~\ref{sec:ch:methodology}). 
Table~\ref{tab:challenges} summarizes this non-exhaustive list of challenges.

\subsection{Systems Challenges}
\label{sec:challenges:system}
\label{sec:ch:system}

\challenge{ch:swdefined}{\refpr{pr:software}}{Make ecosystems fully software-defined, and cope with legacy and partially software-defined systems.}

The scale, diversity, and dynamicity of ecosystems advocates for self-managed control\footnote{Although full self-management is not entirely possible, minimizing the human administrator intervention is key for achieving performance.}.
The largest datacenters in the world span over millions of square feet\footnote{\url{https://www.racksolutions.com/news/data-center-news/top-10-largest-data-centers-world/}}, contain up to hundreds of thousands of compute servers, and tens of thousands of switches and networking equipment. They service up to millions of customers and their diverse workloads. Manually configuring and managing this volume of computing machinery and workloads is infeasible. Herein lies the need of fully \emph{software-defined ecosystems}.  

The key principle behind software-defined ecosystems is the dissociation (i.e., the separation of concerns) between the physical resources and mechanisms, and the software-related interfaces and policies exposed to the users.
Cloud computing has enabled software-defined systems by first virtualizing compute hardware, via virtual machines. In the early to mid 2010s, more resources and services have been virtualized: software-defined networking~\cite{DBLP:conf/hpdc/GangulyABF06,DBLP:conf/sigcomm/JainKMOPSVWZZZHSV13}, software-defined storage~\cite{DBLP:conf/sosp/ThereskaBOKRTBZ13}, and even software-defined security~\cite{DBLP:conf/ndss/ShinPYFGT13}.

The next step towards software-defined ecosystems is the design and implementation of software-defined datacenters~\cite{DBLP:conf/ficloud/DarabsehAJBVR15} or clouds~\cite{DBLP:journals/fgcs/JararwehADBVR16}. The aim is to enable seamless and efficient, possibly federated, composition of software-defined ecosystems. In this paradigm, users and systems developers need not be concerned with low-level hardware configurations and interactions, but rather declare and dynamically change their non-functional requirements: security and privacy policies (e.g., who can access what), level of fault-tolerance (e.g., on how many datacenters must data be replicated), service-level agreements of network performance (e.g., guaranteed bandwidth or latency), scalability, and even trade-offs between availability and consistency.

An important challenge of fully software-defined ecosystems is the integration with \emph{legacy} systems, i.e., partially software-defined. This is an endemic problem in (distributed) computer systems development, as re-designing and re-building successful legacy systems is an inefficient and intricate endeavor. Such problems have been successfully tackled in grid Computing by using an additional layer of indirection, such as a meta-middleware~\cite{DBLP:journals/sj/KerteszK09,DBLP:journals/computer/BalMNDKPKSJ10} that reconciles many different sub-components and brokers their inter-operation.

\challenge{ch:nfr}{{\bf P\ref{pr:nfr}, P\ref{pr:artcrft}}}{Make non-functional requirements first-class considerations, understand key trade-offs between them, and enable ways to specify targets (dynamically) with minimal (specialist) input.}

Customer workloads are increasingly more diverse in terms of volume, variety, velocity, etc., and ultimately of {\it vicissitude}~\cite{DBLP:conf/ccgrid/GhitCHHEI14}, that is, how each of these challenges becomes more prominent at seemingly arbitrary moments of time. To express this diversity, ecosystem customers and operators must agree not only on functional requirements (what to run), but also on increasingly more sophisticated non-functional requirements (NFRs, see also P\ref{pr:artcrft}) and Quality of Service (QoS) guarantees expressed as Service-Level Agreements (SLAs). For example, expressing elasticity could use any or a subset of the over ten available metrics~\cite{DBLP:journals/corr/HerbstKOKEIK16}. When more resources and services will be software-defined~(C\ref{ch:swdefined}), NFRs could include additional SLA terms that relate to how resources and services are used. 

This calls for non-functional requirements (NFRs) to become first-class considerations in the design and operation of ecosystems.

The current practice is to define NFRs for entire applications, including highly reconfigurable applications such as workflows. This can lead to resource waste~\cite{shen2015availability} and inability to express sophisticated needs.
We envision that NFRs could become much more fine-grained than currently in practice. Specifically, we envision \textit{spatial fine-grained NFRs}, that is, expressing detailed NFRs for each unit of work (e.g., task of a bag-of-task, function of a FaaS workflow, microservice of a service-based application), and \textit{temporal fine-grained NFRs}, that is, expressing NFRs that change over time possibly dynamically (i.e., at runtime).

Although finer-grained NFRs can be beneficial, they also increase complexity for the user. 
Understanding and selecting complex NFRs and selecting the right SLOs to meet them can become overwhelming~\cite{dillon2010cloud}.

This raises issues of balancing performance with the usability of cloud platforms.
For example, to enable relatively unsophisticated clients to obtain good performance with minimal (cost) overhead, we envision new methods to automatically translate minimal specialist input into detailed requirements and, consequently, actions taken by the system to adapt. 
Even for expert users, changing NFRs at runtime can be cumbersome.
We further envision that NFRs will change dynamically, to respond to the monitored, predicted, or detected state of both ecosystem and application; for example, changing SLOs upon detecting a resource overload or a straggling task.

A variety of techniques for self-adaptation already exist in the space of cloud computing~\cite{DBLP:books/sp/17/IosupZMKMSAMB17}, including many auto-scaling approaches that work well in practice~\cite{journal/tompecs/IlyushkinAHPEI18}. 
However, much more remains to be done, including: 
(i) investigating the ability of existing formalisms for workflows to express fine-grained NFRs, and designing (parts of) formalisms that can address the missing elements, 
(ii) developing a resource management and scheduling architecture supporting dynamic NFRs as first-class considerations,
(iii) applying NFRs to new elements besides traditional performance, such as exploring the trade-off between power-consumption and other NFRs.

\challenge{ch:heterogeneity}{\refpr{pr:rms}}{Manage extreme heterogeneity.}

Large-scale computer ecosystems exhibit unprecedented, extreme heterogeneity, which we characterize mainly as (i) workload heterogenity, (ii) infrastructure, and (iii) peopleware (addressed in C\ref{ch:sociallyaware}). We discuss these in turn, then formulate the main challenge in this context.

We see workload heterogeneity as (i) functional, that is, applications require special hardware, such as GPUs, and special software, such as FaaS platforms, to function, and 
(ii) non-functional, that is, applications must satisfy SLAs, for example, web applications have low-latency requirements~\cite{brutlag2008user}, whereas large data processing applications are primarily concerned with throughput~\cite{DBLP:conf/ideas/LiuIX14}. Workloads achieve heterogeneity also through (iii) the interplay in the same workload between mixtures of applications with different functional and non-functional requirements.

Corresponding to an increasing workload heterogeneity, we see an increase in infrastructure heterogeneity. GPUs~\cite{DBLP:journals/corr/ChenLLLWWXXZZ15} and TPUs~\cite{DBLP:conf/isca/JouppiYPPABBBBB17} (ASICs) are increasingly used for machine learning applications, and FPGAs are increasingly used for internal datacenter and cloud operations\footnote{\url{https://aws.amazon.com/ec2/instance-types/f1/}}. New kinds of memory, such as Intel Optane (3D XPoint), with its unique latency and throughput characteristics, are now becoming mainstream~\footnote{{https://www.anandtech.com/show/12136/the-intel-optane-ssd-900p-480gb-review}}. Even disregarding these new developments, a plethora of compute and storage types are servicing a variety of needs in computer ecosystems; for example, AWS alone has over 70 types of compute instances (excluding deprecations), and hundreds of cloud services such as Container Service and Lambda, each providing additional options for running compute jobs. This is different from the past, when datacenters were filled with similar hardware, for ease of use and maintenance.

The interplay of heterogeneity in both applications and infrastructure creates new research challenges. How to program applications easily? How to exploit the heterogeneity of infrastructure for optimal performance, cost savings, energy efficiency, etc.? How to control the trade-off between efficiency, and various non-functional and functional requirements? How to create a uniform system out of several heterogeneous components that can be investigated also theoretically?

There is already some work in this area, albeit preliminary. It includes languages and tools to program once and run on heterogeneous hardware~\cite{DBLP:journals/corr/abs-1709-06416}, which extends the large body of previous work on middleware to abstract storage devices and services.

\challenge{ch:sociallyaware}{\refpr{pr:rms}}{Socially aware systems, with the human in the control loop.}

At the scale of computer ecosystems, we observe {\it social hereogeneity}: it is likely that the presence of many distinct individuals can be understood and managed through concepts of social networking, of users creating collective patterns of usage.
Thus, the ``convergence of technological and social networks''~\cite{DBLP:journals/internet/DattaDHI12}, if understood and managed, creates opportunities for better system design and also for better quality of experience for ecosystem-users. 
Preliminary work hints that understanding social-interaction patterns leads to better understanding of resource usage~\cite{DBLP:conf/sigcomm/RoyZBPS15}, to designing more efficient ecosystems~\cite{DBLP:journals/internet/IosupBSJK14}, and to improved application experience~\cite{DBLP:journals/tkdd/JiaSBIKE15}.

We see three main challenges in this context:
(i) understanding, modeling, and predicting the key social interactions, 
(ii) leveraging the models and predictors to improve performance and service-experience, and 
(iii) exploring the trade-off between the degree of control afforded to users, including privacy of their usage data, and the performance the system can achieve. In this work, we explore related work for the first two challenges; for the third, we point to the extensive vision on trust proposed by Epema et al.~\cite{DBLP:journals/ccr/LopezMEDHIBFR15}. 

Seminal work has focused so far on understanding the social relationships between users where the relationships are implicit~\cite{DBLP:journals/tweb/WilsonSPZ12,DBLP:journals/internet/IosupBSJK14}, such as direct communication, direct exchange of data, and acting together. For example, these have been shown to forming strong social relationships (ties) between users of online games~\cite{DBLP:journals/tkdd/JiaSBIKE15}, peer-to-peer file-sharing~\cite{DBLP:journals/tpds/IamnitchiRSF11}, high-performance workloads~\cite{DBLP:conf/hpdc/AamnitchiDG06,DBLP:conf/europar/ZhengSZJF11}, etc. The key issue is to generalize these findings, across different applications, types of ties and of graphs, and method of collecting relevant data.

Using implicit social relationships in computer ecosystems does not need to start from scratch. Our system for collaborative downloads The 2fast~\cite{DBLP:conf/p2p/GarbackiIES06} socially aware protocol leads to optimal data sharing in peer-to-peer networks. Automatic identification of dominant users~\cite{DBLP:conf/grid/IosupDELW06} and of job groupings~\cite{DBLP:conf/europar/IosupJSE07} in scientific grid workloads led to pioneering work by IBM~\cite{DBLP:conf/europar/ZhengSZJF11}. These and more recent studies~\cite{DBLP:journals/internet/IamnitchiBK12,DBLP:journals/tpds/IamnitchiRSF11} indicate that new workload patterns do emerge from implicit social interaction and can be leveraged.

\challenge{ch:adaptation}{\refpr{pr:rms}}{Make use of adaptation approaches, from simple feedback loops to self-awareness, to respond automatically to anomalies and to changes in requirements.}

Adaptation approaches, up to and including self-awareness, can greatly help with a variety of \ourdomainshort{} problems. In our 2017 survey of the field~\cite{DBLP:books/sp/17/IosupZMKMSAMB17}, we have identified 10 classes of such problems with immediate practical use: 
(i) recovery planning, 
(ii) autoscaling of resources, 
(iii) runtime architectural reconfiguration and load balancing,
(iv) fault-tolerance in distributed systems,
(v) energy-proportionality and energy-efficient operation,
(vi) workload prediction,
(vii) performance isolation,
(viii) diagnosis and troubleshooting,
(ix) discovery of application topology, and
(x) intrusion detection and prevention. 

In the same survey~\cite{DBLP:books/sp/17/IosupZMKMSAMB17}, we have also identified 7 classes of existing approaches:
(i) feedback control-based techniques,
(ii) metric optimization with constraints,
(iii) machine learning-based techniques,
(iv) portfolio scheduling,
(v) self-aware architecture reconfiguration,
(vi) stochastic performance models,
(vii) other approaches. 
For each class, we have provided a set of specific problems where the approach has been applied in practice. 

The key remaining challenges are 
to enable adaptation under sophisticated non-functional requirements (challenge identified as part of C\ref{ch:nfr})
and to
(i) select from these approaches those most promising to adapt easily to the expanded scope of entire computer ecosystems, the former as much as possible automatically and the latter with minimal portability effort,
(ii) challenge the existing assumptions that make adaptive methods tractable, e.g., can we complement traditional data structures and algorithms with approaches that are adaptive, deterministic but complex, such as machine-learning-based indexes~\cite{DBLP:journals/corr/abs-1712-01208} and approximate indexes~\cite{DBLP:journals/corr/abs-1801-10207}?
(iii) understand systematically the interplay between different adaptive approaches operating simultaneously or even in conjunction in the computer ecosystem,
(iv) conduct relevant field studies, of putting the adaptive techniques in practice in (near-)production settings,
(v) extend traditional adaptive techniques to include with feedback from ecosystem engineers.

We have for years conducted inroads into these challenges when applying adaptive approaches to resource management and scheduling for datacenter ecosystems~\cite{DBLP:conf/ccgrid/GhitCHHEI14,DBLP:conf/sigmetrics/GhitYIE14,DBLP:journals/computer/BeekDHHI15} (for many more remaining challenges, see Section~\ref{sec:use:dcmgmt}). 

\challenge{ch:scheduling}{{\bf P\ref{pr:rms}, P\ref{pr:artcrft}}}{Scheduling, consisting of both provisioning and allocation, on behalf of different, possibly delegating stakeholders.}

Two phenomena concur to make scheduling in ecosystems more challenging than in typical systems, e.g., in operating systems. First, scheduling occurs on behalf of users on resources typically offered by an operator; this means that the scheduling process must both allocate resources to individual jobs (as in traditional operating and distributed systems), and also provision resources on behalf of the user across super-distributed ecosystems (see P\ref{pr:artcrft})---this is the {\it dual problem} of scheduling in \ourdomainshort{}. 
Second, for many computer ecosystems, new conditions and requirements have appeared. In particular, the diversity of users and the rapid addition of new technologies means workloads can change drastically over both short and long periods of time. For example, grid workloads exhibit short-term burstiness~\cite{DBLP:journals/tpds/Li10} and also increased fragmentation into smaller tasks over long periods of time~\cite{DBLP:journals/internet/IosupE11}. Similar phenomena appear in cloud environments~\cite{bharathi2008characterization, reiss2012heterogeneity}.

We envision a set of new scheduling challenges in this context. 
The dual scheduling problem requires either tight collaboration between users and operators, or partial to full automation of at least the work of one of the sides. 
Users and operators can agree on the use of auto-scalers~\cite{journal/tompecs/IlyushkinAHPEI18}, which provision resources on behalf of users, dynamically, with minimal configuration and expert-level support. Offloading, that is, sending a part of the workload for execution to other resources (and possibly other operators), can also be a useful technique for the user-operator collaboration~\cite{DBLP:conf/ccgrid/OlteanuTI13}.
Users can have their work automated by advanced, typically job-specific, execution engines (e.g., glide-in technology in grids, and the Hadoop execution engine for MapReduce processing depicted in Figure~\ref{fig:bigdata:ecosystem}).
Operators can leverage auto-tiering and multi-level auto-scalers. 
Among the larger set of challenges here, the key remaining challenge in autoscaling consists of 
(i) selecting a good autoscaler that matches the needs of the current workload\footnote{\url{https://atlarge-research.com/lfdversluis/2017-11-24_lfdversluis_autoscaling-comparison.pdf}}~\cite{journal/tompecs/IlyushkinAHPEI18}, possibly dynamically, 
(ii) inventing new autoscalers for emerging workload-characteristics and -needs, and 
(iii) replacing as much as possible the workload-specific with workload-agnostic techniques~\cite{journal/tompecs/IlyushkinAHPEI18}.

The new conditions require adaptation of every traditional approach, e.g., regular scheduling, scavenging for resources, migrating jobs. 
Allocating workloads to the provisioned resources has been a topic of research in regular scheduling for decades, with hundreds of approaches and policies~\cite{jennings2015resource}, but selecting and adapting results to emerging workloads remains non-trivial.
Memory scavenging is a method applied to reduce compute resource consumption, e.g., for scientific workloads~\cite{uta2016towards}. By using small portions of available memory from other tenants or nodes, a relative small performance overhead can be traded for significant gains in resource consumption.
Extending this technique, e.g., for use in a multi-tenant virtual machine environment, can prove to be beneficial for performance and resource consumption, but could lead to conflicts in meeting other SLOs, such as performance isolation or operational risks defined by the SPEC RG group in~\cite{DBLP:journals/corr/HerbstKOKEIK16}.

\challenge{ch:service}{\refpr{pr:rms}}{Sophisticated components in the ecosystem offered as services.}

The emergence of cloud computing is transforming the global ICT industry, from which it employs a significant fraction of skilled personnel and for which it delivers a sizable financial revenue~\cite{tr:idc14cloud}. 

We are entering a period of XaaS (Everything-as-a-Service), where any product or technology can be supplied as a service and delivered to the consumer through the Internet~\cite{DBLP:conf/IEEEcloud/DuanFZSNH15}. 
Enterprises such as Google, Microsoft, Amazon, Oracle, Adobe are moving away from the traditional perpetual license model and are choosing to offer their products on-demand; a consumption-based model. 

Many companies and organizations have transitioned (parts of) their operation to cloud-based services, shifting their operation model to take advantage of the elasticity, availability, security, and pay-as-you-go pricing model of the cloud~\cite{DBLP:journals/fgcs/JararwehADBVR16}; for these companies and organizations, cloud-based services are supplanting in-house infrastructure and legacy software\footnote{\url{https://www.strategyand.pwc.com/reports/zero-infrastructure}}.

The XaaS ecosystem is rapidly expanding, with emerging 'aaS models appearing next to the three main supporting pillars~\cite{mell2011nist}: Infrastructure-as-a-Service (IaaS), Platform-as-a-Service (PaaS), and Software-as-a-Service (SaaS). 
These new service models span one or more of the aforementioned pillars, and include Data-as-a-Service, Benchmarking-as-a-Service, Function-as-a-Service, Authentication-as-a-Service, etc.
This raises interesting new challenges, that relate traditional non-functional requirements to cost, resource waste, and manageability; for example, we discuss some of the challenges of serverless and FaaS operation in Section~\ref{sec:use:apps}.

One of the main challenges of these new 'aaS models is the need for standardization~\cite{DBLP:conf/cloud/Abu-LibdehPW10,DBLP:conf/fedcsis/LabesSRKZ12}. 
The cloud architecture is still subject to rapid change, with hundreds of new technologies being added in each of the past five years. 
Providers each have their own technology stack, raising concerns of vendor lock-in and lack of interoperability. 
A necessary key challenge therefore is to define appropriate reference architectures. 
Moreover, this challenge goes beyond the merely technical: 
it requires communication and collaboration between major cloud operators, and possibly also with their main clients, to ensure adoption.

Currently, all the services in the XaaS ecosystems are offered through the Internet, making the Internet a bottleneck, albeit widely distributed and fault-tolerant. 
Because network performance is crucial, it would be advantageous to have a unified platform that encapsulates both network and cloud services. 
Research is underway to address this issue, for example, through EU-funded projects such as Scalable and Adaptive Internet Solution (SAIL\footnote{\url{http://www.sail-project.eu/}}) and Unify\footnote{\url{http://www.fp7-unify.eu/}}.

\challenge{ch:ecosystem}
{{\bf P\ref{pr:software}--\ref{pr:artcrft}}}
{The Ecosystem Navigation challenge: solving problems of comparison, selection, composition, replacement, and adaptation of components (and assemblies) on behalf of the user.}

Computer ecosystems can seem daunting to starting users. They pose numerous challenges related to the use of complex systems, complex code-bases, and seeming lack of control. For the user who wants to achieve some goal through the use of existing existing, the presence of many open-source components for own deployment and API-based hosted by cloud operators raises the problem of selection and configuration (adaptation). For example, which of the tens of machine instances provided by Amazon EC2 should a researcher start to use? And which of the seemingly similar machine instances to use, among the many available clouds?
For the starting developer, the problems extend to the full set of comparison, selection, composition, etc. This raises {\it the Ecosystem Navigation challenge}, of solving this set of problems on user's behalf, subject to custom requirements.

We envision two main challenges, derived from the \SwEng{} and \DistribSys{} communities, respectively: (i) satisficing the Ecosystem Navigation challenge for the restricted set of ecosystems whose components are all based on explicit, narrow, well-defined APIs, and (ii) satisficing the Ecosystem Navigation challenge for the general case where ecosystem components can use any interface or API, and in particular the ecosystem can include legacy systems and systems with poorly specified interfaces. 

The challenge expressed at (i) has been a hot topic of research in \SwEng{} for the past decade. The early results focus on functional composition of components, with tens of methods already put in practice~\cite{DBLP:journals/tse/RobillardBKMR13}. 
Many challenges remain to be solved, either for the case where the components are hosted and their APIs only minimally specified~\cite{DBLP:conf/icse/WitternYZLDYS17}. As an early example,  IBM~API~Harmony~\cite{DBLP:journals/ibmrd/WitternMLVSRJN16} focuses on automatically analyzing APIs produced by different developers and producing recommendations of components that may be used together.
An open challenge still arises when the composition must also guarantee non-functional requirements are met (so, beyond functional requirements).

The latter remains largely an open challenge, with preliminary work focusing on empirical findings, e.g., performance studies across the entire community trying different combinations of components in the ecosystem. However, comparing performance studies and finding their shared findings even in a narrow field remains difficult. We see as a promising avenue toward addressing this issue the creation of community-wide, general reference architectures, of the kind depicted by Figure~\ref{fig:bigdata:ecosystem}; such a reference architecture can guide the exploration of alternatives of similar nature, but for general systems instead of the components considered for (i).

\challenge{ch:interoperation}{{\bf P\ref{pr:rms}, P\ref{pr:artcrft}}}{Interoperate assemblies, dynamically: geo-distributed, federated, multi-DC operation, and service delegation.}

The digital economy is ever expanding, leading to massive businesses, such as computing services (e.g., cloud services), telecom, media and entertainment providers. Inherently, these service-providers serve a geographically distributed market, with clients spread across the globe. Consequently, datacenters are built close to customers and services delegated (see also P\ref{pr:artcrft}), to ensure good quality of service (e.g., low-latency) and to avoid moving large amounts of data over vast distances. For example, some of the largest datacenter operators manage tens to hundreds of datacenters each\footnote{\url{https://www.datamation.com/data-center/data-center-companies.html}} and some enterprises use more than a single cloud provider\footnote{According to the 2014 EC-commissioned study~\cite{tr:idc14cloud} and to the 2018 IHS Technology report discussed by SDXCentral \url{https://www.sdxcentral.com/articles/news/ovh-takes-aws-azure-google-us/2017/09/}}.

We envision the need for many and eventually all \ourdomainshort{} to operate over multiple, federated, and geo-distributed (micro-)datacenters. We also advocate for effortless composition of geo-distributed ecosystems or services. Such needs stem from several reasons. First, enabling efficient \emph{wide-area analytics}~\cite{DBLP:conf/nsdi/RabkinASPF14} is key for business interoperability: multiple entities that operate in different regions may need to perform analytics over vast collections of geo-distributed data, to derive new insights and to produce value from their joint data. Second, avoiding \emph{vendor lock-in}~\cite{DBLP:conf/cloud/Abu-LibdehPW10} is key for reducing and even optimizing operational costs, limiting the data loss in case of natural disasters, and avoiding the loss of privacy when one vendor is compromised. Third, consolidating the hardware resources of distributed datacenters in a cloud-of-clouds~\cite{DBLP:journals/ibmrd/RochwergerBLGNLMWECBEG09,campbell2009open} improves overall resource utilization, aids in meeting user service-level agreements, and reduces management costs.

The main challenge in achieving interoperability and effortless composition is derived from the core of \DistribSys{}: achieving efficient, secure, and lightweight communication between possibly untrusted systems. 
We envision a community-wide research effort that (i) assesses the feasibility of existing systems in this area, such as grpc\footnote{\url{https://grpc.io/}}, Apache Thrift\footnote{\url{https://thrift.apache.org/}}, or the older Grid communication and interoperability layers, such as Ibis~\cite{DBLP:journals/computer/BalMNDKPKSJ10}, or GridFTP~\cite{DBLP:conf/sc/AllcockBKL05}, (ii) starting from the found limitations, provides the needed extension of existing and older research efforts, and possibly designs radically new communication libraries, and (iii) considers {\it co-design} of communication protocols/libraries and the data analytics engine, such that computation is performed directly on encrypted data~\cite{DBLP:conf/icfec/MakkesUDBB17}, without analyzing in the clear and exposing data on compromised (or malicious) sites.

\subsection{Peopleware Challenges}
\label{sec:challenges:peopleware}
\label{sec:ch:peopleware}

\challenge{ch:engagement}{\refpr{pr:educating}}{Create communities and environments for people to engage with the design and operation of ecosystems.}

Making the core concepts of computer ecosystems accessible to a wide audience is vital for both the society and the continued evolution of \ourdomainshort{}. Currently, many of the concepts involved in modern ecosystems, from heterogeneous datacenters to abstract operational policies, are complex and hard to grasp. This poses barriers for people to engage in the study, design, and operation of computer ecosystems. To address this issue, we see as key issue the creation of appealing and understandable visual and textual abstractions that lower the barrier of entry and facilitate a visual understanding of the dynamic processes typical to computer ecosystems.
We also envision that global competitions in key areas of computer ecosystems (e.g., resource management and scheduling in datacenters, see P\ref{pr:rms}) can encourage engagement with these tools.

The key challenge is to find ``the right model'', that is, to choose the layer of abstraction and the visual/textual domain-specific language, while addressing multiple stakeholders with different levels of sophistication and different problems to explore, and with the simulator still delivering good performance.

One of our contributions to this effort is the OpenDC platform for datacenter simulation~\cite{opendc}. Its visual interface allows users to build their own virtual datacenters and run workloads on their simulated resources, seeing how datacenters operate from an inside perspective. But this concept of a visual builder does not need to be restricted to physical models: we also envision users will create their own scheduling policies in OpenDC (see Section~\ref{sec:use:dcmgmt}). 

We have already used OpenDC in our own research and in the classroom, including in a periodic workshop we give at Restart.network, an education network for refugees in the Netherlands. Entirely through the visual interface of OpenDC, students use OpenDC to build their own datacenters, and to simulate workloads running on them using various policies. 
Among the remaining challenges, we envision here the development of a library of components, assemblies, and workloads, to be shared across the \ourdomainshort{} domain.

\challenge{ch:curriculum}{\refpr{pr:educating}}{Create a teachable common body of knowledge for \ourdomainshort{} (BOKMCS).}

We see as a long-term and perhaps unrealizable challenge the design of an BOKMCS. However, we consider the process of refining the new needs of a curriculum on \ourdomainshort{} as valuable for both the community, and for future students and trainees who will want to join the community. Like Simon~\cite[p.113]{book/Simon96}, we see the value of a general education in the fundamentals of natural sciences or of engineering, and preferably both at a good level. 

Derived from our practical experience with students from various universities and technical universities consistently ranked at the top-level of the global academic establishment, we see a number of important additions to the common ACM/IEEE Curricula Recommendations for Computer Science (2013) and \SwEng{} (2014)\footnote{\url{https://www.acm.org/education/curricula-recommendations}}, and NSF/IEEE-TCPP Curriculum Initiative on Parallel and Distributed Computing (2012)\footnote{\url{https://grid.cs.gsu.edu/~tcpp/curriculum/?q=home}}: 
(i) General problem-solving techniques, covering many and possibly all of the techniques we describe in Section~\ref{sec:core:thisscience},
(ii) Systems Thinking, including elements of Complex Adaptive Systems and Control Theory (see Section~\ref{sec:core:thisscience}),
(iii) Design Thinking, including the representation and evaluation of designs, and designs with quantitative, qualitative, and even no final goals~\cite[Ch.5-6]{book/Simon96}, and possibly advanced cross-field topics in design~\cite{design:book/Lawson05,design:book/Cross11}. 

We also see specific gaps:
(iv) for students following low-quality \SwEng{} courses, we recommend taking more in-depth classes in the area of ``SE/Requirements Engineering'' (ACM/IEEE, p.178) and possibly also ``HCI/User-Centered Design and Testing'' (ACM/IEEE, p.92), especially focusing on the analysis of {\it non-functional} requirements and design/modeling tools that include {\it realistic and quantitative} aspects, and 
(v) for students following primarily a traditional curriculum, we further see the need for learning the basics of experiment design with software artifacts, of conducting systematic and comprehensive literature surveys, and possibly of conducting user studies from ``HCI/Statistical Methods for HCI'' (ACM/IEEE, p.93).

\challenge{ch:explaining}{{\bf P\ref{pr:rms}, P\ref{pr:educating}}}{Support for showing and explaining the operation of the ecosystem to all stakeholders, continuously.}

Currently, many institutions and individuals rely on the availability and correct operation of digital services.
Typically, the operational details of these services are either fully hidden from or merely opaque to the user.
As new ecosystems are developed, especially by combining services from multiple vendors, the operation of these ecosystems becomes even more difficult to oversee for most stakeholders.
Key to regaining this oversight is adding support for showing and explaining the operation of an ecosystem.
We envision that operators of ecosystems will have a duty, possibly legislated, to continuously and transparently inform stakeholders on a variety of operational properties, including risk (e.g., frequency of outages, impact of security breaches, possibility of data loss), cost (e.g., financial, energy), and legal aspects (e.g., licensing, compliance with local laws).

To this end we identify two main challenges.
First, the {\ourdomainshort} community must define metrics to quantify key operational properties and explain these metrics and their implications for each stakeholder.
Although such metrics exist for some established domains and applications (e.g., elasticity and operational risk for cloud computing~\cite{DBLP:journals/corr/HerbstKOKEIK16}, cost models for cloud providers), it is unclear if and how these metrics translate to ecosystems, and which metrics could be important or informative.
Second, we must develop methods for monitoring, inferring, and predicting an ecosystem's operational metrics, from the dynamic metrics measured possibly only by the most transparent of the ecosystem constituents.

\challenge{ch:design}{{\bf P\ref{pr:educating}, P\ref{pr:personalprivilege}}}{The Design of Design.}

We see design as a major challenge for the field of \ourdomainshort{}: not only good designs are difficult to achieve for the level of complexity posed by computer ecosystems, but also students and later practitioners in the field do not have prior training in Design Thinking and sometimes even Systems Thinking (see also C\ref{ch:curriculum}). 

We envision creating design processes that trade-off the rigor and precision needed to make software ecosystems run, and the creativity and innovation needed to make software ecosystems perform new functions and better.
We need to start almost from scratch, with: 
(i) overall, understanding and creating good design processes, for individuals and for teams of designers, that increase the likelihood of obtaining useful designs (a {\it meta-design challenge}),
(ii) understanding and creating ways to represent designs,
(iii) understanding and creating ways to test and to compare designs. 

Further steps, for example, the design of computer ecosystems with organizations in the loop~\cite[p.154-5]{book/Simon96} (instead of merely individuals), and advanced cross-field topics in design~\cite{design:book/Lawson05,design:book/Cross11} remain long-term, open challenges.

Although not a must for practical reasons, society also benefits when the general public understands the basic principles of design and is able to enjoy them as art\footnote{Similar considerations apply to architecture} (see also P\ref{pr:educating}). This is an open challenge for \ourdomainshort{}.

\subsection{Methodological Challenges}
\label{sec:challenges:methodology}
\label{sec:ch:methodology}

\challenge{ch:reproducible}{\refpr{pr:reproducible}}{Simulation-based calibrated approaches and real-world experimentation with methodology that ensures reproducibility as key instruments for problem exploration and solving, and for evaluating and comparing ecosystems.}

The \ourdomainshort{} must create its methodologies and instruments, and show they  can lead to successful problem exploration and solving, and to adequate evaluation and comparison of ecosystems. The consequences of failing in this task are dire: 
meaningful problems cannot be solved, and unreproducible results and ecosystems that are not useful cannot be distinguished from valuable contributions. 
We conjecture that the \ourdomainshort{} community can derive useful methodologies and instruments from two main approaches characteristic to empirical research, each with important benefits and specific challenges: (i) simulation of ecosystems, e.g., through a discrete-event model, and (ii) real-world experimentation based on methodology that leads to reproducibility (see also C\ref{ch:reproresults}). we discuss these approaches, in turn.

Simulation-based experimentation can be fast and scalable relatively to real-world experimentation. Simulation obviates the need for physical access to the resources being simulated, and is thus democratizing research in the field by allowing research groups without significant resources to join the community. However, this approach challenges scientists to develop reasonably accurate models of both the topology and the workload being simulated. Validating that this is indeed the case, thus showing that the model is indeed accurate enough, is not only a key scientific challenge, but also a challenge that requires reopening the discussion about valuable contributions in the field (see P\ref{pr:reproducible})---the community must show it values results such as validation studies. Similar challenges apply to real-world experimentation.

Real-world experimentation has the advantage that the system under test is real, revealing possibly hidden issues that have not been considered by models or whose models are inaccurate. However, real-world experimentation for computer ecosystems has important, pragmatic limitations. The scale of the system under test is typically reduced, and especially large-scale systems on-par with real deployments are rarely available in practice; notable exceptions such as Grid'5000 and the DAS~\cite{DBLP:journals/computer/BalELNRSSW16} are still only medium-scale infrastructure. Moreover, test suites and benchmarks typically cannot take control over the entire system under observation, due to user-access limitations, and their execution time is constrained because it matches that of the real-world system.
There are many technical and research challenges that need to be addressed in real-world experimentation, related to the trade-off between (i) realism and statistical power of experiments, and (ii) duration, cost, and access-rights.

As an exemplary challenge, we discuss event-based modeling and simulation of datacenters. Although many simulators exist~\cite{iosup2008digisim,cloudsim2011,kohne2014federatedcloudsim}, important features such performance variability are only now being added to simulators~\cite{DBLP:journals/simpra/MathaRP17} and being validated. Moreover, applying the existing simulation tools to the diverse set of scenarios we observe in real-world setups is not always a straightforward process. From a too-narrow set of modeled scenarios, to very limited validation with real-world results, we still see important challenges standing in the way of their wider application as system evaluators. Our own work, the open-source simulation platform OpenDC~\cite{opendc}, emphasizes the need for a common, validated basis for datacenter simulation (in conjunction with other features, see C\ref{ch:engagement}).

\challenge{ch:reproresults}{\refpr{pr:reproducible}}{Reproducibility of analysis results regarding functional and non-functional properties of systems, including through a new generation of evolving benchmarks, and through processes and instruments for preserving and sharing benchmarking results.}

Reproducibility is a key concern for \ourdomainshort{}, which follows the concerns raised by large-scale systems~\cite{peng2011reproducible, ince2012case, howe2012virtual, vitek2012r3}.
Reproducing arbitrary experiments, to test claims or to compare with previous approaches, 
is non-trivial.
Many factors influence experiments, besides the system under test and its possibly hidden parameters, including but not limited to the workload, the environment, and metrics.

We see two main directions to explore towards solutions for reproducibility: (i) developing real-world benchmarks that offer a good degree of control, and (ii) ensuring reproducibility through methodological considerations.

It remains an open challenge to build high quality benchmarks addressing the diverse problems the \ourdomainshort{} field addresses.
For example, we have experienced first-hand the challenge of reproducibility in our work on evaluating the performance, and later with benchmarking, distributed graph-processing systems. Following years of experimentation, during which we have exposed various degrees of performance sensitivity to various factors, we have proposed the LDBC Graphalytics~\cite{DBLP:journals/pvldb/IosupHNHPMCCSAT16} benchmark, which has been adopted by companies and researchers in the field. Central to Graphalytics is the idea of objective comparison between graph-processing platforms by controlling the key parameters, using (i) a comprehensive suite of real-world algorithms, and synthetic and real-world datasets, (ii) an extensive set of metrics to quantify system performance, scalability (we quantify horizontal/vertical and weak/strong scalability), and robustness (we quantify failures and performance variability), and (iii) a renewal process to curate and possibly change the algorithms, datasets, and gathered metrics. 
It is symptomatic that other de-facto standard benchmarks in the field do not have the properties (i)--(iii).

To improve reproducibility of experiments, the SPEC RG Cloud group\footnote{\url{https://research.spec.org/working-groups/rg-cloud.html}} is developing new methodologies for (cloud) experimentation.
These include guidelines on reporting metrics and values, specifying the aspects of the environment that can lead to reproducibility, sharing the software and data artifacts used during experimentation, etc.
Additionally, our group is focusing on tools and instruments to gather valuable (anonymized) real-world and synthetic operational traces, and to provide them alongside software artifacts for benchmarking through artifact-repositories available freely to individuals, industry, and academia.
A prime example of this is our current Grid Workload Archive~\cite{iosup2008grid}, in which we provide several real-world traces and basic tools for analyzing them.

\challenge{ch:testing}{\refpr{pr:reproducible}}{Testing, validation, verification in this new world. Manage the trade-offs between accuracy and time to results.}

Testing and validating for correct behavior in distributed computer systems is a strenuous activity from the perspectives of designers and engineers, and pragmatically due to the large amount of (computing) resources needed for such endeavors~\cite{DBLP:journals/fgcs/RemenskaWVTB13}. This task has many intricate variables, many interactions between the system components, and (too) many possible states. 
For \ourdomainshort{}, where many ecosystems intricately interact with each other, the problem of testing and validation is even more difficult difficult, in particular due to the ``curse of dimensionality'' (that is, the search space increases exponentially with the number of states).

To tackle this problem, we envision a dual approach for testing and validation : (i) at a \emph{micro}-level, and (ii) at a \emph{macro}-level. The former focuses on the small \emph{building blocks}, and possibly their (limited) interactions. Conversely, the latter focuses on interactions at the level of entire \emph{ecosystems}. Similarly to our considerations at C\ref{ch:reproducible}, testing and validation techniques in particular at a macro-level have to take into account trade-offs between time-to-solution, amount of resources used, and the quality and quantity of guarantees they provide.

For the \emph{micro}-level testing, we see as the main challenge selecting and adapting the techniques that are already successfully used in practice. Designing benchmarks using a choke-point analysis~\cite{DBLP:conf/sigmod/ErlingALCGPPB15} could expose performance and functionality issues in key components of a system. Another useful approach is model-checking, where specialized tools exist for checking both the design~\cite{DBLP:conf/forte/RemenskaWTVB14} and the implementation~\cite{DBLP:conf/ershov/VijzelaarVFB14} of (distributed) systems.

We believe that testing at a higher level in the hierarchy, or an entire ecosystem, is still an open research question. Previously, we have identified several research directions for IaaS benchmarking~\cite{DBLP:books/sp/14/IosupPE14}. We envision \emph{experiment compression} (i.e., combining real-world experiments with emulation and simulation) as key to achieving sustainable testing, validation, and benchmarking in \ourdomainshort{}. We also propose evaluating the \emph{short-term dynamics} and \emph{long-term evolution} through periodic testing using judiciously chosen frequencies of repetitions~\cite{DBLP:conf/ccgrid/IosupYE11}.

\challenge{ch:science}{\refpr{pr:evolution}}{Build a science of \ourdomain{}. 
Revisit in the process the principles of \DistribSys, \SwEng, \PerfEng.}

Overall, we see as more important the process of trying to define an independent field of science, than actually seeing \ourdomainshort{} recognized as an independent field of science.
We propose a pragmatic approach to meeting the high threshold conjectured by Denning~\cite{DBLP:journals/cacm/Denning13a} for becoming a field of science. The approach consists of:
(i) studying the novel natural and artificial processes that appear in computer ecosystems (the ``Central Premise'' in Section~\ref{sec:core:mcs}), which we argue are pervasive in the modern digital markets and critical to knowledge-based societies;
(ii) defining a body of knowledge and skills that relate to computer ecosystems, as explained in Section~\ref{sec:core:mcs}, based on sound and far-reaching principles (Section~\ref{sec:core:principles}), starting from the already large existing body of knowledge identified in Section~\ref{sec:core:thisscience};
(iii) experimenting with ecosystems and simulating them, to enable discovery and validation, and meet reproducibility and falsifiability principles (Section~\ref{sec:core:principles}, especially P\ref{pr:reproducible});
(iv) contributing to codifying and teaching the body of knowledge and skills developed at point (ii);
(v) complementing (ii), and jointly with all stakeholders, defining, codifying, and teaching a body of ethics relevant for work in computer ecosystems.
As we explain in Section~\ref{sec:core:otherscience}, other sciences have taken this pragmatic approach.

We ask several important questions that can be the next steps in addressing this challenge. 
Considering the large body of existing related knowledge, what existing laws, theories, and concepts from (classic)~\DistribSys{}, \SwEng{}, and \PerfEng{} still apply or do not apply anymore for \ourdomainshort{}? 
What abstractions can we reuse, e.g., can there be an operating system for massivized computer systems? 
What new abstractions are needed for achieving the \ourdomainshort{} vision? 
Last, can we build a science of \ourdomainshort{} from first-principles?

\challenge{ch:newworld}{\refpr{pr:evolution}}{The New World challenge: understanding and explaining new modes of use, including new, realistic, accurate, yet tractable models of workloads and environments.}

This is the challenge of understanding the fundamental properties of the emerging field of \ourdomainshort{}, as described in Section~\ref{sec:core:mcs} under empirical and phenomenological research, and under formal (analytical) modeling. 

One specific difficulty is the inclusion, in such research and models, of {\it versions} and {\it configurations} of the software under study (see C\ref{ch:reproresults}). For this challenge, overall results must be detailed for each set of versions and configurations that diverge significantly from the default software, where significant divergences are to be established both empirically and through validated models.

\challenge{ch:ethics}{\refpr{pr:ethics}}{Understand challenges in the ethics of \ourdomainshort{}, and evolve our instruments to support ethics in this context.}
We envision, non-exhaustively, some of the main issues \ourdomainshort{} should address, and discuss them in turn. First, \ourdomainshort{} should involve all stakeholders and agree on a set of ethical challenges that the community can pragmatically hope to solve.

\ourdomainshort{} must consider the ethics of exclusion and of inclusion afforded by \ourdomainshort{} technology. 
Complex technology can exclude from the use of critical ICT services the significant fraction of the population who lacks expertise, and may even discriminate implicitly~\cite{boyd2014networked}.
So far, the society has started to address {\it algorithmic} bias\footnote{\url{https://www.aclu.org/blog/privacy-technology/surveillance-technologies/new-york-city-takes-algorithmic-discrimination}}, but should the ecosystems running the algorithms also be considered?
At the other end of the expertise spectrum, the openness of the technology allows anyone with a grasp of \DistribSys{} to develop their own ecosystem-components, but also opens up a Pandora's Box of poorly designed, poorly implemented systems, which raise issues including security and privacy. 
Questions arise of transparency (when and how?), potential for bias and abuse (how to assess? how to reduce?), etc.

\ourdomainshort{} must consider the ethics of technology-facilitated anti-social and destructive behavior.
For example, due to its scale, reach, and degree of automation, the Facebook platform is now having to answer hard questions~\cite{issues:Chakrabarti18} about political interference on behalf of various powers (including state operators), false news, social separatism through echo chambers, harassment based on personal views including political. What are the limits of responsibility on the side of the ecosystem scientists, designers, engineers, and organizations and societies where they conduct their work?

The ethics of ecosystem operation in various kinds of markets are also concerns for \ourdomainshort{}.
In free markets, if the history of economics is used as predictor, the current accumulation of technology and skill in the hands of a few large organizations is detrimental to the society~\cite{issues:Economist18a}. In response, anti-trust laws affect already Amazon, Apple, Google, Facebook, Microsoft, and Qualcomm. Their decade-long legal battles with the European Commission over anti-trust laws have intensified in 2018\footnote{\url{https://www.nytimes.com/2018/01/24/business/eu-qualcomm-fine-antitrust.html}}~\cite{issues:Economist18a,issues:Economist18b}, but often the legal battles lack the technological knowledge offered by predictive tools. (Anti-trust decisions may push ecosystems to further open up and, in reverse, to break up main components~\cite{issues:Economist18b}, leading to new ecosystem concerns.)  
In closed markets, state monopolies, and broad legal and executive powers in the hands of state agencies, can lead to large computer ecosystems that reduce civil liberties and human rights in all countries\footnote{\url{https://www.theatlantic.com/international/archive/2018/02/china-surveillance/552203/}}~\cite{journals/ssoar/IntronaW04}.
This raises a meta-question of ethics, What are the limits of market and political aspects our community should consider?, whose answer can define how \ourdomainshort{} will address these issues.

Ethical issues arise also in the operation of our own community, in particular about ethical peopleware. 
Similarly to other sciences, and especially their empirical domains\footnote{``Meticulous record keeping is at the heart of good science, and this is especially true for field scientists and naturalists.''~\cite[Kindle Loc. 114]{concepts:book/Canfield11}.}, we envision that ecosystems science, design, and engineering will document more the key choices, data, and even daily operations. 
We must address the ethics of a publish-or-perish publication in science, which incentivizes low-quality results, citation games, and even copying or ripping off the scientific results of others~\cite{DBLP:journals/cacm/Vardi11b}. Rethinking the publication process~\cite{DBLP:journals/dagstuhl-reports/MehlhornVH12}, the meaning, role, and structure of conferences~\cite{DBLP:journals/cacm/Fortnow09} and workshops~\cite{DBLP:journals/cacm/Vardi11} is within the scope of \ourdomainshort{}.
We must train peopleware, especially designers and engineers, to avoid developing the kind of technology that led in the past couple of years to various class-action lawsuits against large technology companies, e.g., against Uber\footnote{\url{https://arstechnica.com/tech-policy/2017/04/uber-said-to-use-sophisticated-software-to-defraud-drivers-passengers/}}. Universities are taking note of this pressing concern and are starting to provide ethics courses for computer science\footnote{\url{https://www.nytimes.com/2018/02/12/business/computer-science-ethics-courses.html}}, but there is still more to be done.
What are the emerging ethical issues inside our community? How to reduce their occurrence or even avoid them entirely?

%% file: content-main-usecases.tex
\section{Massivizing Computer Systems: Use Cases}
\label{sec:use}

In this section, we discuss application domains and use-cases of \ourdomainshort{} (Sections~\ref{sec:use}.1--6); we also identify classes of applications that will not benefit immediately from advances in \ourdomainshort{} (Section~\ref{sec:use:not}). 
We envision that computer ecosystems built on the principles of \ourdomainshort{} will lead to significant benefits over the current approaches, and in some cases to technology disruption: 
achieving economies of scale (e.g., reducing resource waste and cost), 
ensuring better and more diverse non-functional properties of systems, 
lowering the barrier of expertise needed for use, 
removing the most tedious tasks from the daily tasks of engineers, etc.
This can have immediate impact in many application domains.

\begin{table}[!t]\centering
\ra{1.1}
\begin{tabular}{@{}lll@{}} \toprule
Loc. & Description & Key aspects \\ 
\midrule
& \multicolumn{2}{l}{{\it Endogenous applications}} \\
\midrule
$\S$\ref{sec:use:dcmgmt} & Datacenter management &  RM\&S, XaaS, ref.archi. \\
$\S$\ref{sec:use:apps} & Emerging application structures &  serverless \ourdomainshort{}\\
$\S$\ref{sec:use:graph} & Generalized graph processing &  full \ourdomainshort{} challenges\\
\midrule
& \multicolumn{2}{l}{{\it Exogenous applications}} \\
\midrule
$\S$\ref{sec:use:science} & Future science  & e-, democratized science \\
$\S$\ref{sec:use:onlinegaming} & Online gaming  &  multi-functional \ourdomainshort{}\\
$\S$\ref{sec:use:banking} & Future banking &  regulated \ourdomainshort{}\\
\bottomrule
\phantom{abc}\\
\end{tabular}
\caption{Selected use-cases for \ourdomainshort{}.}
\label{tab:applications}
\end{table}

Table~\ref{tab:applications} summarizes the non-exhaustive list of six application domains we discuss in this work.
We distinguish two directions of application: (i) {\it endogenous}, that is, the computer science and in particular the computer systems areas using the concepts and technologies developed within the science of \ourdomainshort{}, and (ii) {\it exogenous}, that is, domains of application that use ICT and in particular computer systems technology to augment or expand their capabilities. 
Among the endogenous application domains, we count cloud computing and big data as directly benefiting from advances in \ourdomainshort{},
and 
datacenter management (Section~\ref{sec:use:dcmgmt}),  
future application structures (Section~\ref{sec:use:apps}), and 
generalized graph processing (Section~\ref{sec:use:graph})
as application domains of \ourdomainshort{} techniques that enhance and extend existing capabilities. 
Among the exogenous application domains, we foresee the mutual benefits of \ourdomainshort{} and 
e-Science and other forms of computational sciences that use ICT as a core part of their instrumentation (Section~\ref{sec:use:science}), 
online gaming (Section~\ref{sec:use:onlinegaming}), 
banking (Section~\ref{sec:use:banking}), 
etc.

\subsection{Datacenters: Managing the Digital Factories of the Knowledge Economy}
\label{sec:use:dcmgmt}

In the Digital Economy, datacenters 
serve the role of modern factories, producing efficient, dependable services.
Their clients range from scientists running complex simulations and data processing pipelines (further explored in Section~\ref{sec:use:science}), to consumers playing online games and meta-gaming (further explored in Section~\ref{sec:use:onlinegaming}). 
To achieve their promise of efficiency and flexibility, datacenters must both use their resources near-optimally, and cover a broad range of scales and designs: from the large multi-cluster deployments typical to IaaS clouds such as Amazon EC2 and Microsoft Azure, to the cloud-edge~\cite{DBLP:journals/computer/Satyanarayanan17} micro-datacenters more typical to video transcoding and streaming~\cite{DBLP:journals/computer/Ananthanarayanan17}.
This raises numerous scientific, designerly~\cite{design:book/Cross11}, and engineering challenges.

In our previous work~\cite{opendc}, we have identified as a key aim forming efficient and controllable datacenter ecosystems (see C\ref{ch:ecosystems}) and, as a key challenge a fully automated resource management and scheduling system for datacenters, able to address: (i) the core principles of \ourdomainshort{}, and (ii) in particular the challenges introduced in Section~\ref{sec:core:system}.
We address here several of these issues.

We envision datacenters are increasingly equipped with a (fully) software-defined stack (C\ref{ch:swdefined}), managing nearly automatically workload, infrastructure, and peopleware heterogeneity (C\ref{ch:heterogeneity} and C\ref{ch:sociallyaware}). This will alleviate the need for live-teams of engineers spending time on relatively trivial decisions, and instead allow them to focus on 
(i) monitoring, diagnosing, and controlling new and particularly complex workflows and dataflows on behalf of users,
(ii) navigating the ecosystem (C\ref{ch:ecosystem}),
(iii) designing ecosystems, constructing and exploring what-if scenarios, etc. 

\begin{figure}[!t]
\centering
	\includegraphics[width=0.5\textwidth]{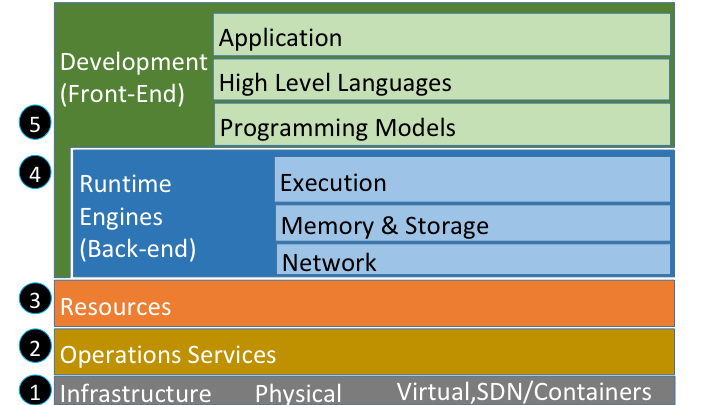}
	\caption{Reference architecture for datacenters (2 levels of depth).}
	\label{fig:refarchidc}
\end{figure}

Datacenters will support increasingly more sophisticated non-functional requirements (C\ref{ch:nfr}), emerging in \ourdomainshort{}, e.g., super-scalability and super-flexibility (see P\ref{pr:artcrft}), or in specialized classes of datacenters, e.g., trust and personalized control in edge-centric (micro-)datacenters~\cite{DBLP:journals/ccr/LopezMEDHIBFR15}.
We envision a guiding, non-mandatory reference architecture for datacenter ecosystems to capture and help manage the diversity of offered services and underlying software layers. For example, Figure~\ref{fig:refarchidc} depicts our reference architecture for datacenters, comprised of 5 core layers, 
{\it Front-end} for the application-level functionality,
{\it Back-end} for task, resource, and service management on behalf of the application,
{\it Resources} for task, resource, and service management on behalf of the cloud operator,
{\it Operations Service} for basic services that are typically associated with (distributed) operating systems, and
{\it Infrastructure} for managing physical and virtual resources; a 6th layer, {\it DevOps} covers functions essential to operating the datacenter but orthogonal to the service provided to customers, such as monitoring, logging, and benchmarking.
Emphasizing the intense focus of the community on simplifying the development of cloud-based applications, the layers closest to the users are further refined into 3 sub-layers each; the sub-layers {\it High Level Languages}, {\it Programming Models}, and {\it Execution} and {\it Memory \& Storage} engines correspond to the similarly named layers in Figure~\ref{fig:bigdata:ecosystem}. 

One of the key emerging capabilities of datacenters is to support complex services (C\ref{ch:service}) while making nearly optimal use of available resources (C\ref{ch:scheduling}). The dual problem of provisioning and allocation is particularly challenging for the diverse ecosystems active in datacenters. The complex approaches that are currently in use rely on complex operations, configured through relatively simple policies only in key points. 
Inspired by the work of Schopf~\cite{Schopf2004}, who proposed in 2004 a detailed 11-step abstraction for the grid scheduling landscape, 
we envision the formulation of a detailed reference architecture for scheduling in datacenters. In this formulation, scheduling is a multi-stage workflow that covers the set of most common actions in datacenter scheduling, with tasks ranging from filtering resources available to the user to task migration. 
We conjecture that this focus on specific stages in the complex scheduling process can facilitate new and competitive designs, and enables newcomers to more easily understand the common structure of schedulers.

The reference architecture for scheduling in datacenters further enables sharing of  entire scheduling solutions or mere components (C\ref{ch:engagement}). Pursuing this goal, we envision~\cite{opendc} a global competition where participants can design and submit their own schedulers, or even parts of schedulers grafted into library-designs of complete schedulers. After simulating their submissions with standard experiments, we will publish the results and announce the winning scheduler. A competition of this sort could foster innovation in the domain, and inspire students to learn more about the process.

\subsection{Science $\circlearrowright$ \ourdomainshort{}, Virtuous Cycle: The Future of Big Science, Democratic Science, and e-Science}
\label{sec:use:science}

We see the future of science as forming a virtuous cycle with that of technology.
Science is increasingly interwoven with the technology that enables it~\cite[Ch.3, loc.903]{design:book/Arthur09}.
Modern science requires experimentation, observation, and reasoning that are possible only through modern technology, and modern technology increasingly complements its own capabilities with the findings of modern science~\cite[Ch.3]{design:book/Arthur09}. Unsurprisingly, as the scientific experiments become increasingly more ambitious and larger, the sophistication and scale of the computer systems supporting them also increase, in a virtuous cycle. We discuss in this section three scientific drivers for \ourdomainshort{}, in turn, Big Science, Democratic Science, and e-Science.
For each, we see \ourdomainshort{} as disruptive technology.

Big Science\footnote{Massive scientific endeavors working as industrial-scale research\cite[p.1]{hiltzik2016big}: large teams, large scientific apparatus, big budgets.} pushes the limits of current computer ecosystems and raises many of the challenges we raise in Section~\ref{sec:challenges}, e.g., massive projects requiring software-defined everything (C\ref{ch:swdefined}), diverse non-functional properties including efficiency and trust (C\ref{ch:nfr}), various forms of system and peopleware heterogeneity (C\ref{ch:heterogeneity} and \ref{ch:sociallyaware}, respectively), etc. 
For example, large scientific experiments rely on federated infrastructure to perform data collection, filtering, and analytics, and especially their storage and processing infrastructure spans federated, geo-distributed data centers. Possibly the best modern example of Big Science is the Large Hadron Collider~\cite{evans2007large}\footnote{The Large Hadron Collider is a successor of the heroic first project in Big Science, initiated by Ernest O. Lawrence's laboratory based on the cyclotron (a small, simple, Nobel-Prize-winning collider) and established through the extension of the laboratory to include a large number of diverse researchers and especially external visitors, all starting in the early 1930s.}, which recently reached the 200 petabyte milestone\footnote{\url{https://tinyurl.com/CERN200PB}}. Analyzing such volume of data is already strenuous for modern computer systems. However, upcoming Big Science projects are expected to deliver even larger volumes and vicissitudes. Such projects include the Square Kilometer Array~\cite{carilli2004science}, the KM3NeT cubic-kilometer telescope searching for neutrinos, or the new Large Hadron Collider\footnote{\url{https://tinyurl.com/LHC3x}} that is expected to be three times as large as its predecessor. We envision \ourdomainshort{} to be the enabler of the computing technology and infrastructure behind such challenging projects, which should provide a sustainable and efficient data processing and storage layers. 

Democratic science is also a scientific driver for \ourdomainshort{}. 
Recent advances in hardware technology have made the market more accessible than ever: storing one gigabyte of data costs below \$0.05\footnote{\url{https://www.backblaze.com/blog/hard-drive-cost-per-gigabyte/}}, while performing 1 GFLOP costs below \$0.1\footnote{AMD Radeon Vega 64 costs $\sim$\$2,000, for 13.7 TFLOPS (single).}. Consequently, it is cheaper and more efficient today to process large amounts of data and to simulate complex situations, than at any point in the history of the human kind. 
This, through XaaS (C\ref{ch:service}), give unprecedented access to science-grade facilities to an increasing number of small laboratories and research groups around the world, enabling the acceleration of scientific discovery without the large funding or teams specific to Big Science, thus democratizing and simplifying access to (virtual) computing infrastructure. Early proposals, such as OurGrid~\cite{DBLP:journals/grid/CirneBACANM06}, envisioned a single, global collaborative grid environment for small teams. With the hardware resources of today, 
we envision infrastructure for enabling large-scale scientific experimentation and discovery, following \ourdomainshort{} principles. 
This could disrupt the elite echelons of science.

e-Science and \ourdomainshort{} can also form the virtuous cycle: fields such as biology, astronomy, and physics are discovering the benefits, but also the challenges, of cloud computing and big data processing~\cite{marx2013biology}.
New and meaningful knowledge is found though the analysis of existing data, but with increased heterogeneity of users, of workloads, and of resources and services, the computer ecosystems in this field must conquer the vicissitude of different ``V''s posing challenges at different moments in time~\cite{DBLP:conf/ccgrid/GhitCHHEI14}.
Many of the applications that run in clouds are structured as shareable workflows, for example, BLAST~\cite{altschul1990basic} and Epigenomics~\cite{bharathi2008characterization} in bioinformatics, LIGO~\cite{bharathi2008characterization} and Montage~\cite{bharathi2008characterization} in computational astrophysics. However, workflow management across heterogeneous resources and services, and scheduling mixtures of workflows on behalf of diverse users, remain relatively open challenges.
The development of technology regarding Internet-of-Things (IoT) is contributing to the amount of data and sensor worldwide.
Trusted data-collection and -processing pipelines, which are crucial when the number of laboratories involved in processing increases, could leverage ecosystems that use novel trust-ensuring techniques for provenance recording and checking (e.g., the emerging blockchain family of technologies).
Smaller-scale than in Big Science, but nevertheless significant, data- and compute-related appear also here, and must be solved for more heterogeneous clients and with much lower budgets.

\subsection{Online Gaming: Can Small Studios Entertain One Billion People with Near-Zero Up-Front Cost?}
\label{sec:use:onlinegaming}

Over one billion players\footnote{The survey company Newzoo reports over 2.2 billion players in 2017 [\url{https://newzoo.com/insights/articles/newzoo-2017-report-insights-into-the-108-9-billion-global-games-market/}], up from the  1.8 billion players of which 711 millions are active as reported by Intel in 2015 [\url{https://blogs.intel.com/technology/2015/08/the-game-changer/}]. The progress reported here is consistent with the multi-decade-long figures provided by the US Entertainment Software Association, for example in its 2017 report [\url{http://www.theesa.com/wp-content/uploads/2017/04/EF2017_FinalDigital.pdf}]. In the US, over two-thirds of the households have gamers who spend significant time playing games weekly; the average age of the gamer is mature, around 35 years.} and 
over a third of a billion spectators\footnote{
Similarly to competitive sports such as (European) football and the Olympic Games, competitive electronic sports (eSports) are rapidly growing in audience. One of the first eSport events dates back from 1999~\cite{wagner2006scientific}.
Social streaming platforms such as Justin.tv, own3d, and more recently Twitch and Youtube Gaming, have greatly contributed to an increase in popularity and interest. 
Live audiences have soared, with the most popular events attractive 20 million unique viewers in 2014, and over 45 million in 2017~\cite{elder2017esports}.
Globally, the number of viewers has increased from 235 million in 2015, to 385 million in 2017, and it is expected that this number reaches 598 million by 2020~\cite{warman2017esports}.}
are valuable\footnote{Newzoo estimates the global market value to have reached \$109 billion in 2017, a continuous increase from the \$70 billion reported in 2012. 
The eSports spectator-related activities are expected to reach \$1.5 billion by 2020, increasing from \$696 million in 2017~\cite{warman2017esports}.} clients of the gaming industry. 
Gaming is not only the most valuable branch of the entertainment industry in most countries, it is also used in enterprise training and employee-interviews, in various forms of serious gaming including what-if and disaster-scenario analysis, and in (higher) education especially for simulation of scenarios.

\begin{figure}[!t]
\centering
	\includegraphics[width=0.5\textwidth]{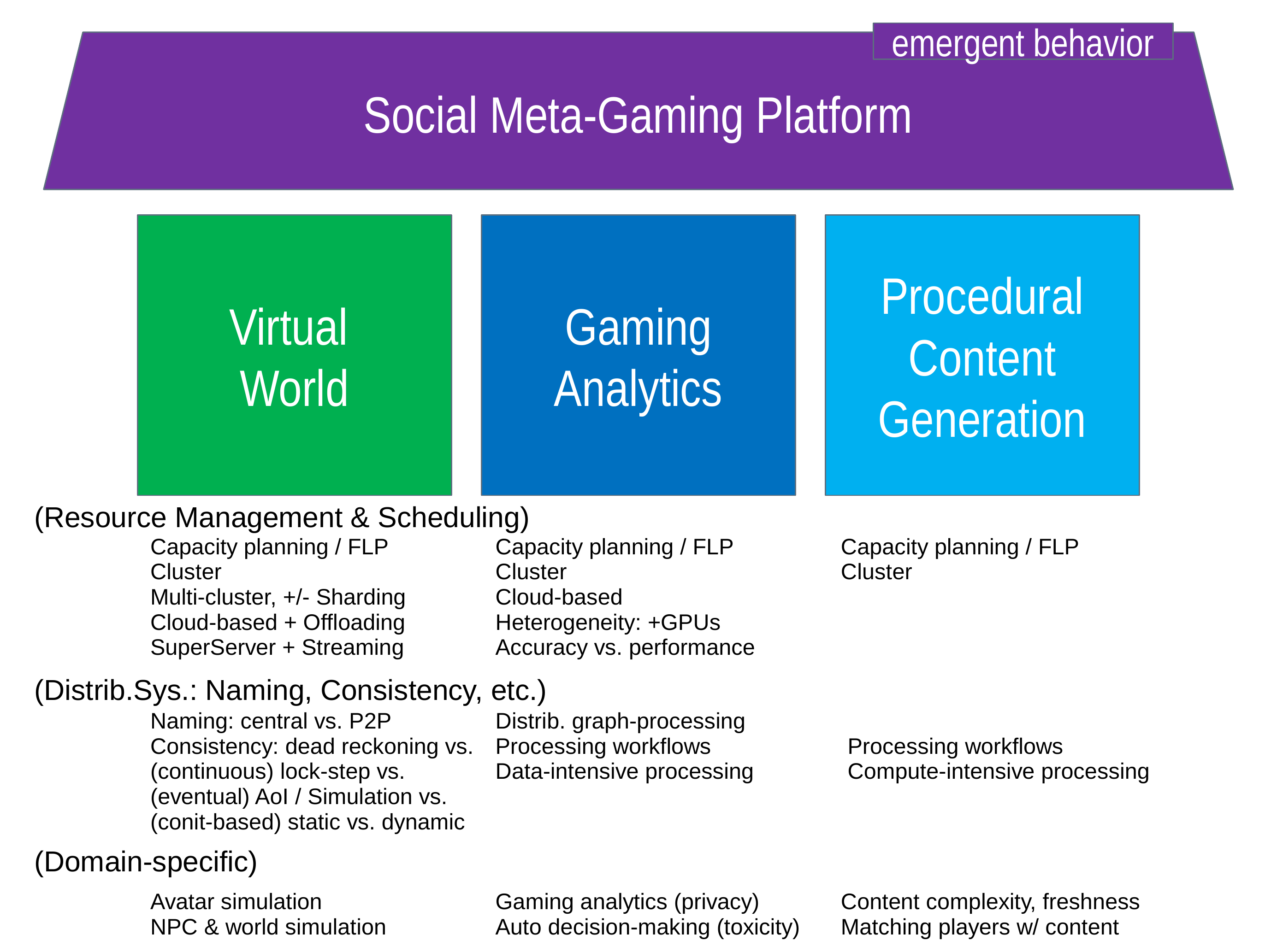}
	\caption{Functional reference architecture for online gaming, with main topics (1 level of depth).}
	\label{fig:onlinegaming}
\end{figure}

Online gaming is a complex, multi-functional application domain. 
Figure~\ref{fig:onlinegaming} summarizes the four key functions present in online gaming through a house-like metaphor.
Besides (i) the game itself, which provides the service of maintaining a seamless {\it Virtual World}, online gaming must also ensure:
(ii) the analysis of game and especially player data, through a {\it Gaming Analytics} platform that supports possibly complex business and operational decisions;
(iii) the generation, curation, and provision of content, from which the automated part is ensured by a platform for {\it Procedural Content Generation}~\cite{DBLP:journals/tomccap/HendrikxMVI13};
(iv) through a 
{\it Social Meta-Gaming} platform, the management and fostering of a community interested in using the game as a symbol relating to diverse, possibly non-game related, activities.

Although, according to the ESA, over half of the frequent players are engaged in multiplayer games through online gaming services, the services they receive remain sub-par: 
(i) the virtual worlds are not seamless, in that they cannot host more than a few thousands of players in the same contiguous virtual-space, and in fast-paced games it is rarely possibly to engage more than a few tens of simultaneous players,
(ii) the player activity is rarely analyzed in depth, correlating social-network and other data across large groups of players is not offered as a service to players, and large teams of community-managers still have to take many decisions case-by-case,
(iii) the game content is rarely updated, rarely player-customized, and never fresh at the scale of the community, and
(iv) the social platform enabling meta-gaming~\cite{conf/MMVE/ShenI11}, that is, spending time in activities related to the game itself, such as playing in a tournament or being spectators, offers only basic tools beyond viewing and sharing of basic content.
These problems stem from deficient gaming ecosystems: the predominant industry approach towards offering online gaming is self-hosting, that is, buying large-scale infrastructure and operating services in-house. This approach does not allow small studios to join the market: the barrier of expertise and of start-up costs is too high. Figure~\ref{fig:onlinegaming} lists many challenges to overcome, grouped by function. We discuss challenges (i) and (ii) in the following; the computational challenge of (iii) is still largely unsophisticated~\cite{DBLP:journals/concurrency/Iosup11} and challenge (iv) still has to overcome first the issues described in Section~\ref{sec:use:graph}. overall, we see online gaming and \ourdomainshort{} as mutually reinforcing, with the solutions provided by \ourdomainshort{} triggering continuously new needs from online gaming.

For the {\it Virtual World}, \ourdomainshort{} disrupts the current approach by promising to eliminate the barriers of entry through the use of third-party services in diverse ecosystems. Can small studios entertain up to one billion people with near-zero up-front costs? By leveraging cloud computing techniques, online games can be massivized~\cite{DBLP:conf/ccgrid/ShenIE13}: they can be positioned close to each player~\cite{DBLP:journals/tpds/NaeIP11} yet cost-effectively~\cite{DBLP:conf/wosp/NaePIF11}, can elastically scale with the ups and downs of active players~\cite{DBLP:conf/europar/ShenDIE13}, and can be made highly available for a fraction of the cost~\cite{DBLP:conf/ccgrid/ShenIICRE15}. Coupled with many other advances from \DistribSys{}, \SwEng{}, and \PerfEng{}, and their inclusion in the larger ecosystem through \ourdomainshort{} techniques, we envision this could solve the key {\it Virtual World} challenges.

For {\it Gaming Analytics}, the challenge of processing connected data at scale remains largely open (see Section~\ref{sec:use:graph}), but the richest and most tech-savvy in the industry have started to leverage data-processing ecosystems. 
Since 2014 and increasingly, the largest gaming companies have started to use third-party data-science services for gaming analytics, e.g., the engineering team at Twitch detailed the use of RedShift for more than 100k queries per second on terabytes of data without manual optimization\footnote{twitch.tv tech blog, \url{https://tinyurl.com/TwitchDataArchi16}}, Blizzard Entertainment has hired Teradata to warehouse data for World of Warcraft, Overwatch, and other popular games\footnote{Since around 2015, see \url{https://tinyurl.com/BlizzardTeradata15}}, and Riot Games has hired Databricks to process data for League of Legends using Spark\footnote{Since 2016, see \url{https://tinyurl.com/RiotDatabricks16}.}
The challenge of enabling this scale and complexity, under a cost model affordable for small studios remains open.

\subsection{The Future of Banking}
\label{sec:use:banking}

Banking is a vital component of modern industries, especially for knowledge-based societies: banking facilitates transferring, depositing, and lending capital. As a heavily regulated industry, banking has been traditionally slow to uptake novel technology and often operates with multi-decade legacy ICT systems. 
However, since 2008 the industry has seen a significant change, combining two contrary directions: (i) more regulation in terms of increased liability and lower tolerance for risk, with (ii) increased openness of the market aiming to provide better service for (retail-)consumers. For example, new regulation appeared in the banking and financial industry, in response to the 2007-2008 financial crisis, including the Basel series of stress-tests. 
To open the market, since 2008 a Single Euro Payments Area (SEPA) helps harmonizing bank transactions in Europe.
To further open the market, in 2015 but with effective implementation in 2018, the European Union has passed into law the second Payment Services Directive (PSD2), which opens up the retail-banking market for more service providers (e.g., Mint for account information), fintech companies (e.g., Adyen and Klarna for payments, Tink for budget management, OurCrowd for crowdfunding), and even the traditional consumer-facing brands (e.g., Google, Apple, telcos, who can combine online retail with banking functionality).
We thus see the future of banking as becoming increasingly dependent on complex ecosystems, and thus requiring the capabilities promised by \ourdomainshort{} while offering back increasingly more dynamic requirements.

We will present our vision through a concrete example. The 
PSD2 regulation is disruptive\footnote{The Economist, An earthquake in European banking, Mar 2017. \url{https://tinyurl.com/TheEconomist17Banking}}, because banks have to open up payment functionality through APIs to other financial operators, and give access to personal data to customers; not only this can lead to banks losing access to their customers (to consumer-facing brands) and lucrative value-adding operations (to fintech companies), but banks are now forced to integrate into a much more complex software ecosystem. 
Moreover, PSD2 enforces strict performance targets, including deadlines in clearing financial transactions such as payments, contracts, and salaries; and offer more customer rights, including the right to refund. 
Security and privacy, which are guaranteed under PSD2, must be reevaluated in the framework of new law, such as the new European General Data Protection Regulation (GDPR)~\cite{albrecht2016gdpr}.
The current banks must address these new challenges, while managing thousands of their own financial applications (over 1,400 at ING, the largest bank in the Netherlands~\cite{DBLP:conf/oopsla/StoelSVB16}) across the diverse and changing legislations of EU's 28 member-states, while facing the significant deficit of skilled personnel that affects the European ICT market\footnote{Europe faces a deficit of over 900,000 IT specialists by 2020~\cite{tr:korte14job}}.

We envision that \ourdomainshort{} can help with the necessary transformations of the emerging banking ecosystems:
(i) by offering banks its core principles (Section~\ref{sec:core:principles}), effectively an ecosystem-oriented framework that allows them to design and compose their software with complex functional and non-functional requirements, and with complex peopleware and methodological issues;
(ii) by building a body of knowledge and skill to approach the daunting problem of combining hundreds of components, some of them legacy applications, others provided by third-parties, into an ecosystem offering meaningful guarantees of its properties;
(iii) by making resource management and scheduling a key building block, capable of ensuring the complex non-functional requirements appearing in this domain, including deadlines, availability, security, and privacy;
(iv) by making self-awareness, and in particular the ability to auto-scale and to tolerate failures (and malicious behavior), a key concern; 
(v) by considering the inherent heterogeneity due to operation in private datacenters, private and public clouds, and edge~\cite{DBLP:journals/ccr/LopezMEDHIBFR15} and fog~\cite{stolfo2012fog} computing infrastructure.

The remaining challenges remain staggering, but we see \ourdomainshort{} as the only alternative to understand and create these complex ecosystems. 
For example, much of banking software must pass strict validation tests, which cannot be easily done; early attempts from \SwEng{}, such as the use of formal specifications that represent banking knowledge and can be validated in Rascal~\cite{DBLP:conf/oopsla/StoelSVB16}, are already useful but still have high \DistribSys{} and \PerfEng{} thresholds to pass: can they validate for such diverse and large ecosystems?, can they validate for non-functional requirements such as performance and availability?
with \PerfEng{}, Petri nets and other workflow tools can help with understanding transition of data, but privacy legislation is still changing, so how can we account for the different views on security and privacy that diverge per country and/or culture~\cite{ion2011home}?

\subsection{The Future of Apps: Serverless, Service- and Workflow-based Ecosystems}
\label{sec:use:apps}

\begin{figure}[!t]
\centering
	\includegraphics[width=0.5\textwidth]{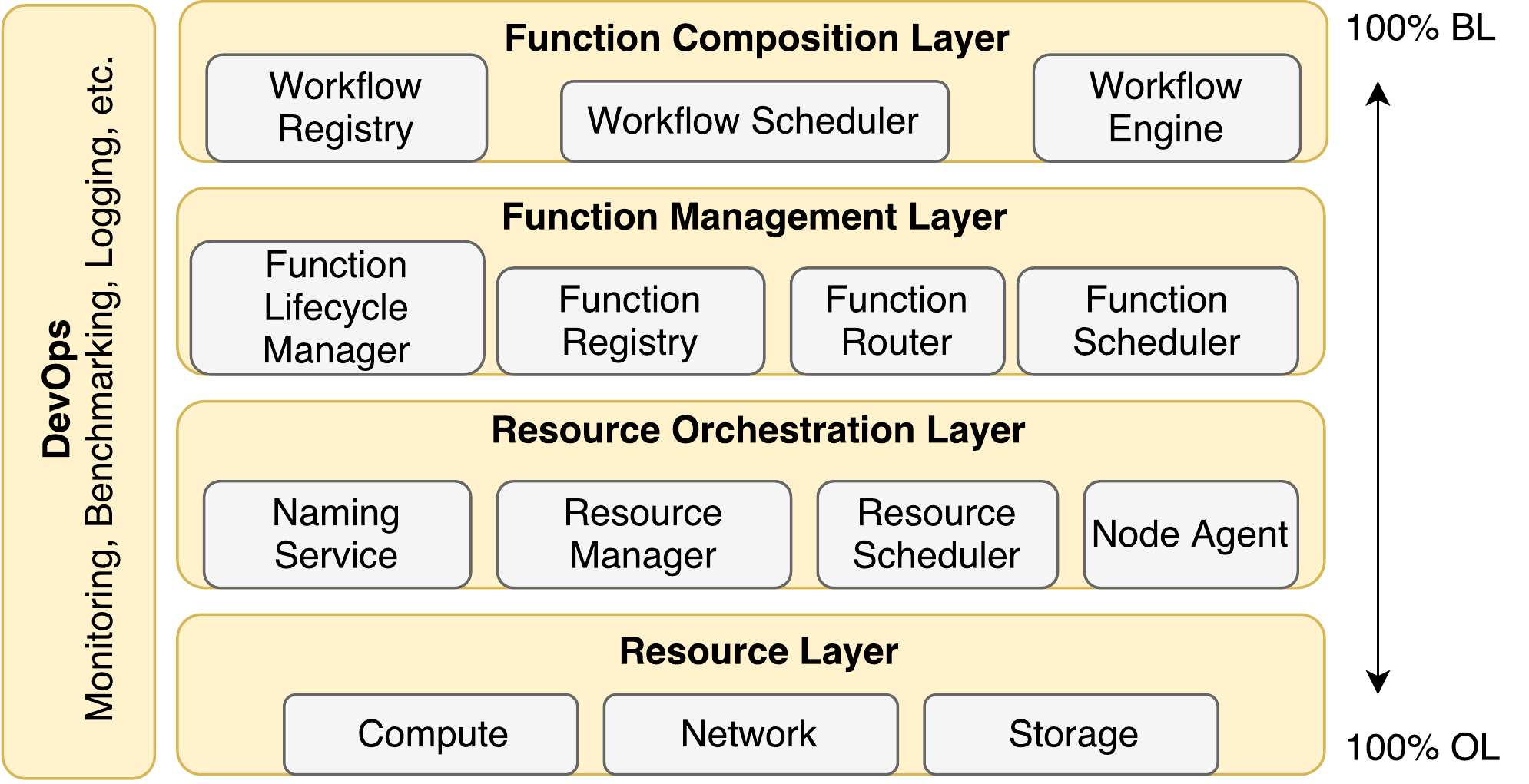}
	\caption{FaaS Reference Architecture ordered from business logic (BL) to operational logic (OL). (Developed jointly with the SPEC RG Cloud group.)}
	\label{fig:faas_refarch}
\end{figure}

Since 2011, starting with grid computing workloads~\cite{DBLP:journals/internet/IosupE11}, we are observing a transition in software-development paradigms and best-practices, from the old, coarse-grained, monolithic projects, to the more modern, fine-grained approach of splitting projects into ever-smaller, independent, service-based components, often referred to as \emph{microservices}. 
As a consequence of this transition, since 2016 cloud vendors are beginning to offer \emph{serverless} computing services; on-demand services billed at a very fine resource-granularity.~\cite{eyk2017spec}
Moreover, within this domain of serverless computing, the \emph{Function-as-a-Service}~(FaaS) paradigm is emerging (see C\ref{ch:service}), offering users the ability to deploy and execute arbitrary functions on cloud infrastructure without the burden of resource management.

User-defined functions are typically stateless and interact with each other through an event-driven paradigm, or through a separate storage layer.
These FaaS workloads can often be modeled as (complex) workflows. 
Therefore, research and engineering efforts will have to address the challenges of workflow and resource management, and of scheduling subject to complex non-functional requirements, in this complex design space.

Because the field is currently in its inception, we see establishing a comprehensive reference architecture for FaaS platforms as a key challenge to overcome.
This will provide the community with a valuable conceptual tool: consistent terminology, a set of typical components, and enough low-level details to focus on the key problems. 
The pragmatic challenges we envision in the FaaS and serverless domains are related to achieving good performance while isolating the operation of each function across multiple tenants. 

Based on the analysis of the major open-source and of a handful of closed-source FaaS platforms so far, we have led the creation, jointly with the SPEC RG Cloud group, of the preliminary reference architecture depicted in Figure~\ref{fig:faas_refarch}. 
Derived from the coarser reference architecture for datacenters depicted in Figure~\ref{fig:refarchidc}, our FaaS reference architecture is based on the notions of business logic derived from a typical serverless application of a FaaS user, i.e., image translation and processing. The {\it Resource Layer} represents the available resources within a cloud. These resources are managed by the {\it Resource Orchestration Layer}, which is often implemented by modern IaaS orchestration services (i.e., Kubernetes) and corresponds to layer 3 in Figure~\ref{fig:refarchidc}. 
On top of these orchestrated resources, the {\it Function Management Layer} manages instances of the cloud-function abstraction, by scheduling and routing functions (the runtime engine in layer 4 in Figure~\ref{fig:refarchidc}); and the {\it Function Composition Layer} is responsible for the meta-scheduling, that is, creating workflows of functions and submitting the individual tasks to the management layer (layer 5 in Figure~\ref{fig:refarchidc}).
To validate this reference architecture, we have already matched its components with real-world FaaS platforms such as OpenWhisk\footnote{\url{https://openwhisk.apache.org/}} and Fission\footnote{\url{https://platform9.com/fission/}}.

\subsection{The Future of Data: Generalized Graph Processing for the Modern Society}
\label{sec:use:graph}

Graphs, and more generally, connected-data\footnote{We define connected-data as unstructured and distributed repositories of data, which can be connected through logical and functional relationships to infer knowledge.} are elegant and powerful abstractions that enable rich analysis. The analysis ranges from traditional graph algorithms (e.g., graph search, shortest paths, transitive closures), to modern machine learning~\cite{DBLP:journals/pvldb/LowGKBGH12} and to deep learning on graph data~\cite{DBLP:conf/cvpr/JainZSS16}. There are many (open) data sources, such as taxi trip data\footnote{\url{https://data.cityofnewyork.us/Transportation/2016-Yellow-Taxi-Trip-Data/k67s-dv2t}}, real-time traffic speed\footnote{\url{https://data.cityofnewyork.us/Transportation/Real-Time-Traffic-Speed-Data/xsat-x5sa}}, historical event repositories\footnote{\url{http://diveplus.frontwise.com}}, social networks\footnote{\url{http://konect.uni-koblenz.de/}}, IoT sensor data\footnote{\url{http://dsa.labs.vu.nl:5001}} (e.g., measuring noise pollution or air quality) that could be used to enrich and improve the lives of the modern society.  

The key challenge here is the lack of an adequate computer ecosystem (the all-encompassing C\ref{ch:ecosystems}): we are still lacking the technologies that can be integrated to connect, explore, query, and analyze such data sources, and further derive knowledge with societal impact. We believe that through \ourdomainshort{} generalized graph-processing would solve this problem by: (i) offering a powerful abstraction for exploring such data repositories and creating links between them, (ii) leveraging all the previous knowledge on how to efficiently process graphs, (iii) taming the necessary computing infrastructure and making it compatible with the demands of both the challenging workloads, and the users. 

As \ourdomainshort{} is solving the basic issues of the generalized graph-processing problem, we also envision that new requirements will emerge, inspired by practical use. 
Current approaches in mining connected-data are only beginning to scratch the surface of the importance of graph processing for the modern society. Recent studies have already shown its applicability by building systems and abstractions to combat human trafficking~\cite{DBLP:conf/semweb/SzekelyKSPSYKNM15}, monitor wildfires~\cite{DBLP:conf/edbt/KoubarakisKM13}, study the human brain~\cite{DBLP:journals/neuroimage/FornitoZB13}, and discover new drugs~\cite{hopkins2008network}. What can \ourdomainshort{} further enable?

\subsection{Which Applications Will Not Benefit?}
\label{sec:use:not}

We also identify application domains that will not benefit immediately from \ourdomainshort{}, among them:
(i) tiny applications with few users or modest requirements for resources, and whose instances work in isolation;
(ii) super-high-performance applications, such as high-frequency trading;
(iii) all the applications that seem contrary to operation in an ecosystem, as described in Section~\ref{sec:core:ecosystems}.

%% file: content-related.tex
\begin{table*}[tb]\centering
\ra{1.3}
\begin{tabular}{@{}lcclcccc@{}}\toprule
 & \multicolumn{2}{c}{Emergence} & \phantom{abc} & \multicolumn{4}{c}{Epistemological Characteristics$^{*}$} \\
 \cmidrule{2-3} \cmidrule{5-8}
Field (Decade Emerging) & Crisis & Continues & & Objectives & Object & Methodology & Character \\ 
\midrule
Modern Ecology (1990s) & Biodiversity loss & Ecology and Evolution & &  DS & Biosphere & ADHS & AC\\
Modern Chem. Process (1990s) & Process complexity & Chemical Engineering & &  DE & Chemical proc. & ADHSP & ACEM\\
Systems Biology (2000s) & Systems complexity & Molecular biology &  &  S & Biological sys. & AHS & ACEMTU\\
Modern Mech. Design (2000s) & Process sustainability & Technical Design & & DE & Mechanical sys. & DHSP & ACEM\\
Modern Optoelectronics (2010s) & Artificial media & Microwave technology & &  S & Metamaterials & DHSP & ACEMTU\\
\midrule
{\bf \ourdomainshort{} (this work)} & {\bf Systems complexity} & {\bf \DistribSys{}} &  &  {\bf DES} & {\bf Ecosystems} & {\bf ADHSP} & {\bf ACES}\\
\bottomrule
\multicolumn{8}{l}{
{\tiny Acronyms follow the framework of Ropohl~\cite[p.4--7]{engineering:Boon06}:
Objectives: D = Design, E = Engineering, S = Scientific. 
Methodology: A = abstraction, D = design (abductive creation), H = hierarchy, I = idealization, S = simulation, P = prototyping.
}}\\ 
\multicolumn{8}{l}{
{\tiny Character: A = applicability, C = approved by the scientific/design/engineering community, E = empirically accurate, H = harmony between results, M = mathematically detailed, S = simplicity, T = truth, U = universality.}}\\
\phantom{abc}\\
\end{tabular}
\caption{Comparison of fields. The row for \ourdomainshort{} is envisioned.}
\label{tab:othersciene}
\end{table*}

\section{Related Work}
\label{sec:related}

In this section, we compare \ourdomainshort{} with other paradigms, first, of computer science and second, of sciences and technical sciences.

\subsection{Is \ourdomainshort{} New?}

We argue that asking ``Is \ourdomainshort{} new?'' is not a well-formulated question.
As any paradigm derived from successful existing paradigms, in this case, the paradigms of \DistribSys{} we discuss in Section~\ref{sec:core:othercompsci}, \ourdomainshort{} faces a question of novelty~\cite[Ch. ``Did we know it all along?'']{biosys:book/AlberghinaW05}. However, deciding the novelty of a field using a vaguely defined notion of novelty can lead to absurd results, such as finding that physics and computer science are not novel fields of science\footnote{{\it Reductio ad absurdum}: We could ask: Is the entire computer science applied mathematics and physics? Are mathematics and physics applied philosophy? And answer: Computer science could be seen as applied mathematics and physics, in that computer science theories and artifacts operate as the artifacts of mathematics and the laws of physics predict.
Mathematics and physics could be seen as applied philosophy, in that they operate in the logical and knowledge framework provided by the latter.
But this is absurd, because we know today that computer science and physics are fields of science. {\it QED}. Arthur gives a more cogent, but book-length argument~\cite{design:book/Arthur09}.}. It is also an open challenge to define a concept of novelty that is not vague, for a field addressing heterogeneous users, workloads, resources and services, processes, etc., as \ourdomainshort{} does.

Instead, we focus on the real thresholds of establishing a new field, introduced by Denning~\cite[p.32]{DBLP:journals/cacm/Denning13a}. We have discussed these thresholds in Section~\ref{sec:core:methodology} and posed the challenge of meeting them gradually in Section~\ref{sec:ch:methodology} (C\ref{ch:science}). 
We address next, in Section~\ref{sec:core:othercompsci}, the key distinguishing elements that currently characterize \ourdomainshort{}.


\subsection{Vs. Other Paradigms Emerging from Distributed Systems}
\label{sec:core:othercompsci}

In the large field of distributed systems, we identify three major paradigms that \ourdomainshort{} builds upon: cluster, grid, and more recently, cloud computing and edge computing~\cite{DBLP:journals/ccr/LopezMEDHIBFR15}.
Much like science is in a constant co-evolution with technology, all these fields have historically co-evolved together with the application types demanded by their users. Cluster computing served the needs of tightly coupled, possibly communication intensive, high-performance scientific computing applications. Grid workloads were mostly high-throughput computing applications, i.e., long-running, conveniently parallel applications~\cite{DBLP:conf/grid/IosupDELW06}, such as bags of tasks or scientific workflows. Together with the data deluge, and the fourth paradigm of data-driven scientific discovery~\cite{hey2009fourth}, analytics workloads, available now in many shapes and formats (e.g., MapReduce, stream processing, machine learning), have migrated to the cloud, which offers a unified platform for cost-efficient computation and storage. 

We have explored in Section~\ref{sec:intro} some of the shortcomings of previous paradigms, such as grid, cloud, peer-to-peer, and edge computing, when addressing challenges of computer ecosystems.
We have also summarized in Section~\ref{sec:intro} and expanded in Section~\ref{sec:bg:problems} five classes of problems that these paradigms have not solved, but are addressed by \ourdomainshort{}. 
Overall, in contrast with these paradigms, \ourdomainshort{} focuses on new problems and challenges (i.e., related to ecosystems, considering peopleware, and the combined spectrum science-engineering-design), for which it offers new views (e.g., ecosystems-first), new and powerful (predictive) concepts and techniques including a synthesis of techniques across \DistribSys{}, \SwEng{}, and \PerfEng{}, and new and existing but improved technologies and instruments (e.g., ecosystems studied {\it in silico} or through full-stack simulation). 

Furthermore, \ourdomainshort{} brings computing closer to the users, empowering them to control how computing ecosystems behave by means of expressive, modern non-functional requirements (such as elasticity and security) and by considering universal access to services to also include less sophisticated users.


\subsection{Parallels with and Other Fields of Science}
\label{sec:core:otherscience}

We see the emergence of \ourdomainshort{} from \DistribSys{} as a process similar to the emergence of other science domains, which we have witnessed in the past three decades. 
Table~\ref{tab:othersciene} summarizes how \ourdomainshort{} matches emergent sub-fields of other science domains, following the framework of Ropohl~\cite[p.4--7]{engineering:Boon06}. Overall, we find that similar goals and approaches as taken by \ourdomainshort{} have emerged from other domains of science and practice. 

These emerging fields have started humbly, part of a broader paradigm. Through useful evolutions of the respective domains of science and practice, they have then developed 
into domains themselves~\cite{DBLP:journals/cacm/Snir11}. This is the model that we envision for \ourdomainshort{}.

Among the fields we survey, closest to \ourdomainshort{} is \BioSys{}. In contrast to \BioSys{}, which has a distinctly scientific orientation and thus a character focused especially toward universality and mathematical formulation of results, \ourdomainshort{} focuses explicitly on design and synthesis (engineering) in its objectives, and on the pragmatic, empirical character of its results. Although we do not exclude a later re-focus of \ourdomainshort{} on universal, mathematically formulated models and theories, for the moment we see a large gap between theory and practice that prevents this development without groundbreaking progress.

%% file: content-conclusion.tex
\section{Conclusion}
\label{sec:conclusion}

Responding to the needs of an increasingly digital and knowledge-based society, we envision ever-larger roles for vast and complex combinations of distributed systems that serve individuals and human-centered organizations.
However, current technology seems ill-equipped to achieve this vision: an ongoing systems crisis hampers not only evolving toward the vision, but even the current operation of modern distributed systems.
In this work, we propose an alternative, \ourdomain{} (\ourdomainshort{}). 
\ourdomainshort{} focuses on systems combined into ecosystems, with scientific, design, and engineering techniques evolving from modern \DistribSys{}, \SwEng{}, and \PerfEng{}, and with a focus on peopleware and methodological outlook beyond mere technology.

Our contribution is five-fold:
(1) we define \ourdomainshort{} to focus on a new central paradigm, computer ecosystems, that distinguishes it among the sub-fields of \DistribSys{}, 
with respect for the rights of expert and non-expert individuals,
and with various elements that we believe can lead organically, in the long-term, to the formation of a new field of science;
(2) we propose a set of core principles for \ourdomainshort{}, including principles that go beyond mere technological aspects, such as scientific, design, and ethical concerns;
(3) we propose a diverse set of challenges focusing on systems, peopleware, and methodological aspects derived from the core principles;
(4) we identify and explore various benefits we envision \ourdomainshort{} can bring to the future of six application domains, both in the area of modern computer systems and to use-cases such as Big Science, democratized science, and e-Science;
(5) we contrast \ourdomainshort{} with both paradigms of modern computer science fields such as \DistribSys{}, and emerging fields of science and technical science.

We have started to address the research agenda formulated in this article, both as a single research group and, through the SPEC RG Cloud group, in collaboration with numerous academic and industry partners. We hope this vision-article will stimulate a larger community to join us in addressing these complex yet rewarding challenges.